\documentclass[aip,jcp,amsmath,amssymb,reprint]{revtex4-1}

\usepackage{graphicx}
\usepackage{color}
\usepackage{framed}
\usepackage{array}
\usepackage{soul}
\usepackage[normalem]{ulem}
\usepackage{caption}
\usepackage{physics}
\usepackage{hyperref}
\usepackage{amsmath}
\usepackage{caption}
\definecolor{gray}{rgb}{0.8,0.8,0.8}
\definecolor{darkgray}{rgb}{0.5,0.5,0.5}
\definecolor{litegray}{rgb}{0.9,0.9,0.9}
\definecolor{shadecolor}{rgb}{0.5,0.5,0.5}

\def\a0{{$a_{\rm 0}$}}

\begin{document}

\title{Influence Functional Approach to Non-Perturbative Exciton Binding Renormalization from Phonons}

\author{Rohit Rana}
\affiliation{Department of Chemistry, University of California, Berkeley, CA, 94720, USA}
\affiliation{\mbox{Materials Sciences Division, Lawrence Berkeley National Laboratory, Berkeley, CA, 94720, USA}}

\author{Eric R. Heller}
\affiliation{Department of Chemistry, University of California, Berkeley, CA, 94720, USA}
\affiliation{\mbox{Chemical Sciences Division, Lawrence Berkeley National Laboratory, Berkeley, CA, 94720, USA}}

\author{Antonios M. Alvertis}
\affiliation{Department of Physics, The University of Texas at Austin, Austin, TX 78712}
\affiliation{Oden Institute for Computational Engineering and Sciences, The University of Texas at Austin, Austin, TX 78712}

\author{Jeffrey B. Neaton}
\affiliation{\mbox{Materials Sciences Division, Lawrence Berkeley National Laboratory, Berkeley, CA, 94720, USA}}
\affiliation{Department of Physics, University of California, Berkeley, CA, 94720, USA}
\affiliation{Kavli Energy NanoScience Institute, Berkeley, CA, 94720, USA}

\author{David T. Limmer}
\email{dlimmer@berkeley.edu}
\affiliation{Department of Chemistry, University of California, Berkeley, CA, 94720, USA}
\affiliation{\mbox{Materials Sciences Division, Lawrence Berkeley National Laboratory, Berkeley, CA, 94720, USA}}
\affiliation{\mbox{Chemical Sciences Division, Lawrence Berkeley National Laboratory, Berkeley, CA, 94720, USA}}
\affiliation{Kavli Energy NanoScience Institute, Berkeley, CA, 94720, USA}

\date{\today}

\begin{abstract}
We construct a many-body model Hamiltonian to capture how phonons renormalize exciton binding as a function of temperature. By using the GW approximation and density functional perturbation theory, we are able to parameterize this Hamiltonian completely from first principles. To capture static quasiparticle properties non-perturbatively, we evolve this Hamiltonian in imaginary time with path integral Monte Carlo using an influence functional based approach. For a class of Wannier--Mott type excitons, our binding energies are in quantitative agreement with experiment. We find that in addition to long-range dipolar interactions from longitudinal optical modes, short-ranged deformation potentials from acoustic modes and transverse optical modes can significantly renormalize electron and hole polaron binding energies at elevated temperature. However, exciton binding energies are only appreciably renormalized by coupling to optical phonons. 
\end{abstract}


\maketitle

\section{Intro}
The evaluation of optical excitation energies in semiconductors is complicated by a myriad of physical factors. Excitations in extended systems are often strongly correlated, forming quasiparticle excitonic states, whose charged distributions couple to the surrounding lattice.\cite{mahan2013many} As a consequence, to achieve accurate estimates of these energies, numerical techniques must be developed that capture the interplay between electronic and nuclear degrees of freedom while maintaining computational efficiency.\cite{dai2024theory} 
In this work, we develop a downfolded Hamiltonian approach to evaluate exciton binding energies using path integral Monte Carlo (PIMC).\cite{ceperley1995path} The Hamiltonian is parameterized by ab initio electronic structure methods, such as density functional theory (DFT) and many-body perturbation theory within the GW approximation\cite{hedin1965new,rohlfing2000electron}, which provide an accurate representation of effective electronic degrees of freedom. An influence functional calculation within PIMC is  used to compute non-perturbative and temperature-dependent exciton binding energies renormalized by coupling to phonons across a variety of polar semiconductors.\cite{feynman2018statistical,feynman2000theory,limmer2024statistical} Using this approach, we find that although acoustic and transverse optical phonons can significantly affect single charge polaron binding energies, they play only a minor role in renormalizing exciton binding compared to longitudinal optical phonons. Our estimates of exciton binding energies are systematically in better agreement than those reported using standard approaches.\cite{bechstedt2005quasiparticle,bokdam2016role}

An exciton is a quasiparticle in which an electron and hole interact electrostatically with each other creating a correlated, long-lived bound state. The exciton binding energy, the reversible work to separate the electron--hole pair, renormalizes the optical gap of a semiconductor, lowering it relative to expectations from the single-particle states forming the material's band structure.
For this reason, significant effort has been directed towards developing an accurate theoretical framework for this excited state problem. The favorable system size scaling of DFT makes it a good candidate for obtaining band energies in solids.\cite{kohn1965self,payne1992iterative} However, DFT is fundamentally limited in its ability to capture many-body electronic correlations, as it cannot, by construction, describe two-particle excitations.\cite{perdew1983physical,sham1983density} The GW approximation provides a correction to the DFT band energies.\cite{hedin1965new,hybertsen1986electron} The Bethe-Salpeter Equation (BSE) approach then builds upon this approximation by including screened electron--hole interactions, typically accounting for electronic screening in the random phase approximation and static limit.\cite{rohlfing2000electron,onida2002electronic} Nonetheless, the standard ab initio GW--BSE method tends to overestimate the magnitude of exciton binding energies in comparison with experiment.\cite{bechstedt2005quasiparticle,schleife2018optical} The main reason for this is that phonons can renormalize the electron-hole interaction.\cite{filip2021phonon,alvertis2024phonon,park2024theoretical,dai2024theory,dai2025polarons} The polarization generated from phonons can interact with the electron and hole, introducing a dynamical, retarded interaction that goes beyond static lattice screening. To account for the charge-phonon interaction in the GW--BSE framework, one can introduce a phonon-screened electron-hole kernel. The limitation of this approach is that the kernel is typically introduced perturbatively.\cite{filip2021phonon,alvertis2024phonon,dai2025polarons}

Path integral methods like PIMC have shown promise in capturing exciton--phonon interactions.\cite{park2022nonlocal,park2022renormalization,park2023biexcitons,rana2025interplay,park2024theoretical} Consequently, computed exciton binding energies and radiative recombination rates from these methods tend to be in good agreement with experiment. However, PIMC is typically applied to simple model Hamiltonians parameterized empirically from experimental measurements of the dielectric constants and effective masses limiting their predictive capability. Here, we present an ab initio framework to parameterize an excitonic Hamiltonian complete with interactions with arbitrary phonon modes. We then exactly integrate out those phonon modes using an imaginary time influence functional that is sampled with PIMC. This allows us to capture non-perturbative static exciton properties as a function of temperature, which for strongly coupled systems like MgO is essential to achieve close agreement with experiment. The resultant zero temperature exciton binding energies validate this approach, and the temperature dependence of these energies reveals the role of thermal phonon screening on exciton stability, allowing us to identify conditions for the onset of free carrier generation.

\section{Theory and Implementation}
We aim to model the energetics of excitons in bulk, three-dimensional semiconductors and how these energies are renormalized by interactions with phonons. To proceed, we consider a Hamiltonian decomposable as
\begin{equation} \label{mostgeneral_short}
    \mathcal{H} = \mathcal{H}_{\text{e}} + \mathcal{H}_{\text{ph}} + \mathcal{H}_{\text{int}}  
\end{equation}
where $\mathcal{H}_{\text{e}}$ represents the Hamiltonian for the electronic degrees of freedom, $ \mathcal{H}_{\text{ph}}$ that for the phonon degrees of freedom, and $\mathcal{H}_{\text{int}}$ the interaction between the two. Generically, the electronic Hamiltonian can be expressed as a sum over bands, $\lambda$, and wavevectors $\mathbf{k}$, as
\begin{multline} \label{mostgeneral_eh}
\mathcal{H}_{\text{e}}  = \sum_{\mathbf{k},\lambda} E_{\lambda,\mathbf{k}} \hat{c}^\dagger_{\lambda,\mathbf{k}} \hat{c}_{\lambda,\mathbf{k}} +\\ \sum_{\mathbf{k},\mathbf{k}',\Delta\mathbf{k},\lambda,\lambda'} U_{\Delta \mathbf{k}}^{\lambda,\lambda'} \hat{c}^\dagger_{\lambda,\mathbf{k}+\Delta \mathbf{k}} \hat{c}^\dagger_{\lambda',\mathbf{k}'-\Delta \mathbf{k}} \hat{c}_{\lambda',\mathbf{k}'} \hat{c}_{\lambda,\mathbf{k}}
\end{multline}
where $E_{\lambda,\mathbf{k}}$ is the energy of the $\lambda$'th band at wavevector $\mathbf{k}$, $U_{\Delta \mathbf{k}}^{\lambda,\lambda'}$ is the interaction between electrons in different bands $\lambda$ and $\lambda'$ with an exchange of momentum $\Delta \mathbf{k}$. The operators $\hat{c}^\dagger_{\lambda,\mathbf{k}}$ and $\hat{c}_{\lambda,\mathbf{k}}$ are creation and annihilation operators for electrons in the $\lambda$'th band at wavevector $\mathbf{k}$, respectively. The basis of single particle plane wave states is appropriate for the class of semiconductors we consider here, which are weakly correlated in their ground state.

While Eq.~\ref{mostgeneral_eh} is general, we will restrict our attention to band edge transitions such that $\lambda$ and $\lambda'$ include only the valence  and conduction bands, $\lambda=\{\mathrm{c},\mathrm{v}\}$. Further, we will consider low energy excitons such that $E_{\lambda,\mathbf{k}}$ is well approximated by an effective mass model, 
\begin{equation}
E_{\lambda,\mathbf{k}} = E_{\lambda}+\frac{\hbar^2 k^2}{2 m_\lambda}
\end{equation}  
where $m_\lambda$ is the effective mass of the valence or conduction band, $\hbar$ is Planck's reduced constant, and $E_{\lambda}$ the energy of band $\lambda$ at the zone center. Since we consider charge neutral systems, we identify the conduction band mass as an electron mass, $m_\mathrm{c}=m_e$, while that for the valence band as the mass of a hole, $m_\mathrm{v}=-m_h$. The interaction matrix element, $U_{\mathbf{k}}^{\lambda,\lambda'}$, is assumed to be Coulombic, 
\begin{equation} \label{coulomb}
U_{\mathbf{k}}^{\lambda,\lambda'}=\frac{e^2}{\epsilon_0 \epsilon_\infty k^2} \left (1-\delta_{\lambda,\lambda'}\right )
\end{equation}  
screened by a single frequency independent dielectric constant $\epsilon_\infty$, where $e$ is the charge of the electron and $\epsilon_0$ is the permittivity of free space. The Kronecker delta $\delta_{\lambda,\lambda'}$ ensures that the interactions are only across the band, between the electron and hole, and attractive. These two simplifications encode a Wannier--Mott exciton model.\cite{knox1963theory,wannier1937structure,mott1938conduction} In the absence of interactions with phonons, this model is exactly solvable yielding hydrogenic bound states.\cite{knox1963theory} 

We assume that the lattice is well approximated by a  collection of phonons that couple linearly with the electronic degrees of freedom. This allows us to write the phonon Hamiltonian as  
\begin{equation} \label{mostgeneral_ph}
\mathcal{H}_{\text{ph}}  = \sum_{n,\mathbf{q}} \hbar\omega_{\mathbf{q}}^n \hat{b}^\dagger_{n,\mathbf{q}} \hat{b}_{n,\mathbf{q}}
\end{equation}
\noindent
where $\hbar\omega_{\mathbf{q}}^n$ is the energy of the $n$'th phonon band at wavevector $\mathbf{q}$, while $\hat{b}^\dagger_{n,\mathbf{q}}$ and $\hat{b}_{n,\mathbf{q}}$ are the associated creation and annihilation operators. We ignore anharmonic effects here that could be encoded with phonon interactions or using effective phonon theories.\cite{zacharias2023anharmonic,lahnsteiner2022anharmonic,park2022nonlocal} Consistent with the effective mass model of the electronic degrees of freedom, we will assume limiting low energy forms for the phonon band structure. For acoustic modes this means assuming a linear dispersion, $\omega_{\mathbf{q}}^n=v_s q$, where $v_s$ is the speed of sound and $q$ is the magnitude of the phonon wavevector. Moreover, optical modes are taken to be dispersionless.

The electron--phonon interaction Hamiltonian for our minimal band model is
\begin{equation} \label{mostgeneral_int}
\begin{split}
\mathcal{H}_{\text{int}} =  \sum_{n,\mathbf{q},\mathbf{k},\lambda} g_{\lambda}^n( \mathbf{q}) \hat{c}^\dagger_{\lambda,\mathbf{k}+\mathbf{q}}\hat{c}_{\lambda,\mathbf{k}}\left( \hat{b}_{n,\mathbf{q}} + \hat{b}^\dagger_{n,-\mathbf{q}} \right)
\end{split}
\end{equation}
which is bilinear in the phonon displacement and band occupancy. The electron--phonon matrix element, $g_{\lambda}^n(\mathbf{q})$, depends on electronic band $\lambda$, phonon momentum $\mathbf{q}$, and phonon mode $n$.  For our Wannier--Mott exciton model, we can associate the coupling to the conduction band with an electron--phonon interaction $g_{\mathrm{e}}^n(\mathbf{q})$ while that to the valence band is associated with a hole-phonon interaction $g_{\mathrm{h}}^n(\mathbf{q})$, which enter into the Hamiltonian with opposite signs due to the particle-hole structure of the exciton.\cite{anselm1956zh,takagahara1999theory} In principle, $g_{\lambda}^n(\mathbf{q})$ depends on the electronic momentum $\mathbf{k}$ as well, but we evaluate these matrix elements at the conduction band minimum and valence band maximum. 

For different phonon modes, there are distinct expectations for the electron--phonon coupling elements in the low wavevector limit. For the acoustic branch of centrosymmetric solids, one can derive the lowest-order coupling in the long-wavelength limit for either the longitudinal (LA) or transverse modes (TA) as\cite{li2021deformation,murphy2018acoustic,lundstrom2000fundamentals}
\begin{equation} \label{acoustic_deform_g}
g_{\lambda}^\text{A}(\mathbf{q}) = D_\lambda^\mathrm{A} \sqrt{\frac{\hbar}{2V \rho v_{s}}} q^{1/2}
\end{equation}
\noindent
where $D_\lambda^A$ is the deformation potential for the $A=\{\mathrm{LA},\mathrm{TA}\}$, $V$ is the volume of the unit cell, and $\rho$ is the mass density of the unit cell. 
For optical phonons, the longitudinal modes (LO) generate a polarization and couple to electrons through a charge-dipole interaction\cite{frohlich1954electrons,feynman2018statistical},
\begin{equation} \label{LO_g}
    g_{\lambda}^\text{LO}(\mathbf{q}) = C^{\text{LO}}_\lambda \sqrt{\frac{\hbar}{2V \rho \omega_\mathrm{LO}}} \frac{1}{{q}}
\end{equation}
where $C^{\text{LO}}_\lambda$ is related to the Fröhlich coupling constant in the limiting case of an isotropic lattice and single active LO mode. The transverse optical modes (TO) do not generate long-range polarizations across the lattice at the dipole level of theory. However, they can still couple with charges through the deformation potential mechanism, which arises from local changes in the crystal potential induced by the lattice displacement pattern. One can show that this coupling at lowest order is constant and given by\cite{li2021deformation,potz1981theory}
\begin{equation} \label{to_deform_g}
g_{\lambda}^\text{TO}(\mathbf{q}) = D_\lambda^{\mathrm{TO}} \sqrt{\frac{\hbar}{2V \rho \omega_\mathrm{TO}}} 
\end{equation}
where $\omega_\mathrm{TO}$ is the transverse optical mode frequency.  These limiting low $q$ forms will serve as the basis of fits for explicitly evaluated matrix elements.

\subsection{Quasiparticle path integral}
To study the system defined by Eq.~\ref{mostgeneral_short}, we employ imaginary time path integrals. The utility of doing so is two-fold. First, it is a natural means to study temperature-dependent quantities as it works directly with the canonical partition function. Second, the bilinear electron--phonon interaction allows us to integrate out the lattice degrees of freedom reducing the complexity of simulations at the cost of introducing an imaginary time influence functional. To proceed, we identify quasiparticles with the electronic degrees of freedom. For the conduction band, we will associate an electron, while for the valence band we will associate a hole. We will express the Hamiltonian in real space, such that the resultant action $\mathcal{S}[\mathbf{x}_e,\mathbf{x}_h,\{\mathbf{u}_n\}]$ depends on the positions of the electron, $\mathbf{x}_e$, and hole $\mathbf{x}_h$, and set of phonon displacements $\{\mathbf{u}_n\}$ for each mode $n$. Due to the form of the Hamiltonian, the action decomposes as
\begin{equation} \label{totAct}
\mathcal{S} = \mathcal{S}_{\text{e}}[\mathbf{x}_e,\mathbf{x}_h] + \mathcal{S}_{\text{ph}}[\{\mathbf{u}_n\}] + \mathcal{S}_{\text{int}}[\mathbf{x}_e,\mathbf{x}_h,\{\mathbf{u}_n\}]
\end{equation}
where $\mathcal{S}_{\text{e}}[\mathbf{x}_e,\mathbf{x}_h]$ is the electronic action, $\mathcal{S}_{\text{ph}}[\{\mathbf{u}_n\}]$ is the action for the phonons, and $ \mathcal{S}_{\text{int}}[\mathbf{x}_e,\mathbf{x}_h,\{\mathbf{u}_n\}]$ encodes the interaction between the two.

The partition function, $\mathcal{Z}$, for the system at fixed temperature, volume and number of particles, is given by an integral over all positions and displacements weighted by the action,
\begin{equation} \label{Partition}
\begin{split}
\mathcal{Z} &=  \int \mathcal{D}[ \boldsymbol{x}_e, \boldsymbol{x}_h,\{\mathbf{u}_n\} ] \exp \left[-\frac{1}{\hbar} \mathcal{S} \right]\\
&=\mathcal{Z}_{\text{ph}} \int \mathcal{D}[ \boldsymbol{x}_e, \boldsymbol{x}_h] \exp \left[-\frac{1}{\hbar} (\mathcal{S}_\mathrm{e}+ \mathcal{S}_\mathrm{nl})\right]
\end{split}
\end{equation}
and can be partially integrated over all of the phonon displacements by virtue of their quadratic energies, leaving an added contribution to the action for the remaining electron and hole, $\mathcal{S}_\mathrm{nl}$, and a constant factor of the noninteracting phonon partition function $\mathcal{Z}_{\text{ph}}$. This is a consequence of the fact that the phonons reduce to a Gaussian field theory.\cite{limmer2024statistical} The remaining statistical weight for the electron and holes depends on the electronic action,
\begin{equation} \label{electronic_H_wan}
   \mathcal{S}_\mathrm{e} = \int_\tau \frac{m_e}{2}\dot{\mathbf{x}}_{e,\tau}^2 + \frac{m_h}{2}\dot{\mathbf{x}}_{h,\tau}^2- \frac{e^2}{4\pi \epsilon_0 \epsilon_{\infty} |{{\bf x}}_{e,\tau} - {{\bf x}}_{h,
   \tau}| }
\end{equation}
which includes kinetic and interaction terms, and the integral over $\tau$ has limits of 0 and $\beta \hbar$ where $\beta^{-1}=k_\mathrm{B} T$ is the temperature times Boltzmann's constant. Additionally, the statistical weight depends on an imaginary-time influence functional $\mathcal{S}_\mathrm{nl}$. The influence functional is nonlocal in imaginary time,
\begin{equation} \label{influence_general_compact}
 \mathcal{S}_{\text{nl}} = -  \sum_{n} \sum_{c,d \: \in [e,h]}\int_{\tau} \int_{\tau '} \int_{\mathbf{q}} \chi_{|\tau-\tau'|}^{n,\mathbf{q}}   F_{cd}^{n,\mathbf{q}}(\mathbf{x}_{c,\tau} -\mathbf{x}_{d,\tau'} )  
\end{equation}
and is expressed as a sum over phonon modes, $n$, electronic bands, $c,d$, two imaginary times $\tau$ and $\tau'$ and a product of a phonon correlation function $\chi_{|\tau-\tau'|}^{n,\mathbf{q}}$ and a kernel $ F_{cd}^{n,\mathbf{q}}(\mathbf{x}_{c,\tau} -\mathbf{x}_{d,\tau'} )$ both of which are most compactly represented in Fourier space with wavevector $\mathbf{q}$. The phonon correlation function has a standard form\cite{feynman2018statistical},
\begin{equation} \label{full_Chi}
    \chi^{n,\mathbf{q}}_{|\tau-\tau'|} = \frac{\cosh[\omega^n_\mathbf{q} (|\tau- \tau^\prime|  - \beta \hbar/2)]}{\sinh(\beta \hbar \omega^n_\mathbf{q}/2)}
\end{equation}
while the kernel is
\begin{equation} \label{int_kernel}
    F_{cd}^{n,\mathbf{q}}(\mathbf{x}) = \frac{V \Gamma_{cd}}{16\pi^3} |g^{n}_{c}(\mathbf{q} )|\cdot|g^{n}_{d}(\mathbf{q} )| e^{i\mathbf{q}\cdot \mathbf{x}}
\end{equation}
where $\Gamma_{cd}$ is 1 for $c = d$ and -1 for $c \ne d$. The influence functional has two clear effects. The self-terms, $c=d$, are strictly negative and describe the attraction of each carrier to its own lattice distortion, corresponding to polaronic effects for the single charges. The cross terms that act between the electron and hole have an interpretation as a frequency and wavevector dependent screening function, attenuating the attraction between the two. Both serve to renormalize the exciton binding energy by stabilizing free charges and screening their interactions.

\begin{figure*}[t]
\includegraphics[scale = 0.376]{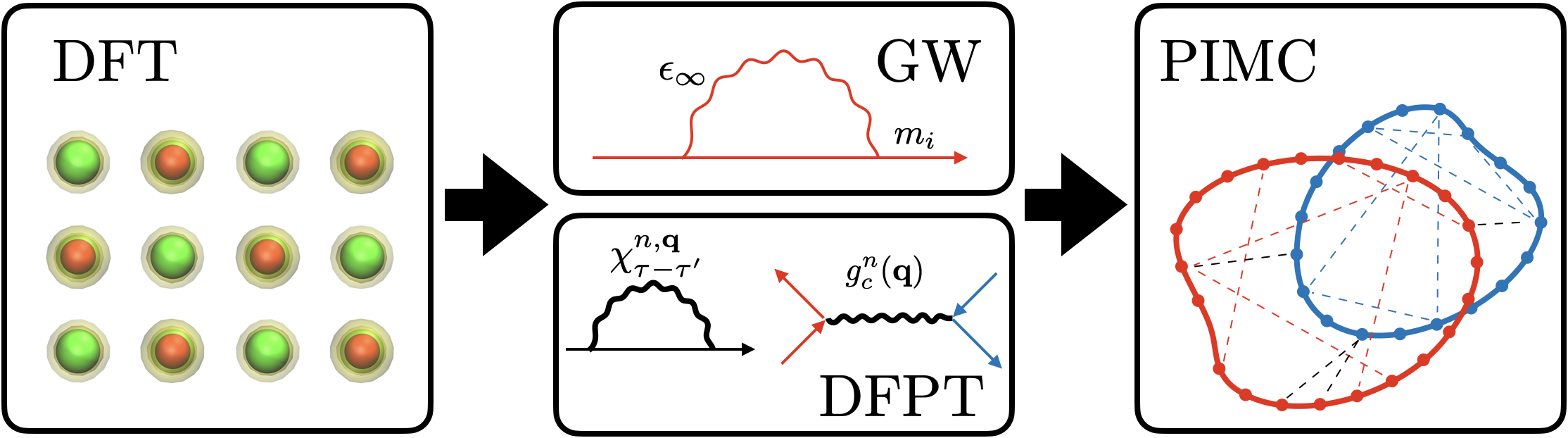}
\caption{Workflow of PIMC framework with first principles parameterization. For the PIMC step, black dashed lines represent the screened Coulomb interaction between the electron and hole from the phonons. Red and blue lines represent the induced self-attraction from the phonons for an electron and hole, respectively.} 
\label{work}
\end{figure*}
We will primarily be concerned with using this path integral formalism to extract thermodynamic properties like the average energy of the system. The average energy can be evaluated from the $\beta$ derivative of the logarithm of the partition function,
\begin{equation}
\langle E \rangle = -\left ( \frac{\partial \ln \mathcal{Z}}{\partial \beta}\right )
\end{equation}
where $\langle \dots \rangle$ denotes an ensemble average. To determine the exciton binding energy, we will introduce ensembles that include both electron and hole in a simulation domain, with averages denoted $\langle \dots \rangle_{eh}$ as well as ensembles with only an electron, $\langle \dots \rangle_e$ or hole $\langle \dots \rangle_h$. The exciton binding energy, $E_\mathrm{B}$, is then computable from $E_\mathrm{B}=\langle E \rangle_{eh}-\langle E \rangle_e-\langle E \rangle_h$ or the difference in the energy of the system when both electron and hole are present and that of the energies of the electron and hole separated infinitely far apart. We will evaluate polaron binding energies, $E_\mathrm{p}$, of an electron or hole from an ensemble in which a single electron or hole is present and all of the electron--phonon coupling terms are set to zero, with an average denoted $\langle \dots \rangle_{i,0}$ for $i=\{e,h\}$. The polaron binding energy of the electron is given by $E_\mathrm{p}=\langle E \rangle_e-\langle E \rangle_{e,0}$ and an analogous expression holds for the hole. Stabilizing polaron binding can thus weaken exciton binding by reducing the free charge energies.

\noindent

\subsection{Numerical sampling of path integrals}
To evaluate ensemble averages computationally, we must discretize the corresponding effective action into $N_{\tau}$ imaginary time slices. From a primitive discretization,\cite{suzuki1976generalized,trotter1959product,ceperley1995path} we can perform standard Monte Carlo simulations, treating the imaginary time slices as particles. The discretized action including the influence functional is
\begin{equation} \label{elec_act_dis}
\begin{split}
    \mathcal{S}_{\text{e}} +&\mathcal{S}_{\text{nl}}= \sum_{t=0}^{N_{\tau}-1} \sum_{c \: \in [e,h]} \frac{m_{c}N_{\tau}}{2 \beta \hbar} (\mathbf{x}_{c,t+1} - \mathbf{x}_{c,t} )^2\\  
    &- \sum_{t=0}^{N_{\tau}-1} \frac{\beta \hbar e^2}{4\pi \epsilon_0 \epsilon_{\infty}N_{\tau} \sqrt{|\mathbf{x}_{e,t} - \mathbf{x}_{h,t}|^2 + r_{\text{c}}^2 }}\\
    &-\frac{\beta^2 \hbar^2}{N_{\tau}^2}  \sum_{n}\sum_{c,d \: \in [e,h]}\sum_{s,t=0}^{N_{\tau}-1} 
    G_{cd}^n(\mathbf{x}_{c,s} -\mathbf{x}_{d,t})
    \end{split}
\end{equation}
where $\mathbf{x}_{i,t}$ is a three-dimensional position vector of quasiparticle $i$ at imaginary time $t \beta \hbar/N_{\tau}$. In order to regularize the Coulomb divergence, $r_{\text{c}}$ is a pseudopotential parameter chosen such that the bandgap of the material is reproduced at 0 separation. As the partition function corresponds to the propagator with the same initial and final positions, we impose periodic boundary conditions in imaginary time, $\mathbf{x}_{i,N_{\tau}} = \mathbf{x}_{i,0}$, thus leading each of the quasiparticles to be represented as ring-polymers.\cite{feynman2018statistical,chandler1981exploiting} The last term in Eq.~\ref{elec_act_dis} results from the influence functional and generates interactions between all pairs of imaginary time slices which is given by
\begin{equation}
G_{cd}^n(\mathbf{x})=\int_{\mathbf{q}}  \chi_{|t-s|\beta \hbar/N_{\tau}}^{n,\mathbf{q}}   F_{cd}^{n,\mathbf{q}}(\mathbf{x})
\end{equation}
a Fourier transform of the imaginary-time discretized phonon correlation function and electron--phonon coupling elements. In general, the integral does not have a closed form. In practice we evaluate it numerically, using a two-dimensional interpolation over imaginary time difference and real space
separation.
\begin{table*}[t]
\caption{\centering Electronic Structure Parameters and Dielectric Constants}
\label{tab:material_parameters}
\begin{tabular}{lcccccccc} 
\hline
System & \: $a$ (\AA) & \: $V $ (\AA$^3$) & \: $\rho $ ($m_0$/\AA$^3$) & \: $E_g$ (eV) & \: $m_e $ ($m_0$) & \: $m_h $ ($m_0$) & \: $\epsilon_\mathrm{s}$ & \: $\epsilon_\infty$ \\  
\hline\hline
MgO       & \: 3.01$^a$          & \: 19.28 & \: 3805.00 & \: 7.95  & \: 0.34 & \: 5.0 & \: 11.3 & \: 3.3 \\ 
CdS   & \: 4.20$^a$              & \: 52.42 & \: 5022.58 & \: 2.50  & \: 0.12 & \: 2.01 & \: 10.4 & \: 6.2  \\
AgCl      & \: 3.98$^a$          & \: 44.43 & \: 5878.32 & \: 2.75  & \: 0.26 & \: 0.85 & \: 15.7 & \: 4.9  \\
CsPbBr$_3$ & \: 8.25$^b$         & \: 796.20 & \: 5308.04 & \: 1.85  & \: 0.20 & \: 0.20 & \: 18.6 & \: 4.4  \\
\hline
\multicolumn{9}{l}{$^a$ Lattice constants obtained from the Materials Project database.\cite{jain2013commentary}} \\
\multicolumn{9}{l}{$^b$ Orthorhombic structure with $a$ = 8.25 \AA, $b$ = 11.75 \AA, $c$ = 8.20 \AA .\cite{linaburg2017cs1,filip2021phonon}} \\
\end{tabular}
\end{table*}

For the PIMC simulations, we initialize each ring polymer via a random walk, where successive beads are displaced by Gaussian-distributed steps with standard deviation $\sigma = \sqrt{\hbar^2/(N_\tau k_\mathrm{B} T m_c)}$, reflecting the free-particle thermal wavelength between adjacent imaginary-time slices. The electron and hole polymers are centered at the simulation box center with an initial separation chosen on the order of the bare exciton Bohr radius $a_X = 4\pi\epsilon_0\epsilon_{\infty} \hbar^2 / \mu e^2$, where $\mu = m_e m_h / (m_e + m_h)$ is the reduced mass of the exciton. Periodic boundary conditions are enforced on all bead positions with a cubic box of length 1000 Bohr. We use a standard Metropolis-Hastings algorithm\cite{metropolis1953equation,hastings1970monte} along with the staging method to sample the action.\cite{tuckerman1993efficient, sprik1985staging,brualla2002path} In staging, we make a coordinate transformation such that we can diagonalize and sample the kinetic part of the electronic action exactly. For each Monte Carlo step, we can propose to move contiguous segments at a time. To improve the convergence and variance of energy estimates, the virial estimator was implemented.\cite{herman1982path,giansanti1988variance} We find $N_{\tau} = 2500$ provides good convergence for energies. Simulations were run for $5 \times 10^4$ Monte Carlo steps, with the first $10^4$ steps discarded for equilibration. Staging move parameters were tuned to maintain acceptance rates of 30–50\%. Furthermore, in order to extrapolate averages to 0 K we used a two-level system fit based on the Boltzmann distribution for a range of temperatures between 50 K and 400 K.

\section{Methods of ab initio calculation}
The basic workflow for the parameterization of the model Hamiltonian and extraction of exciton binding energies is illustrated in Fig.~\ref{work}. The properties needed are extractable from band structure calculations of the electrons and phonons, and evaluation of coupling matrix elements from response theories. Below we provide the details for such calculations for the systems used to study the utility of this approach, which include MgO, CdS, AgCl, and CsPbBr$_3$.

\subsection{Electronic bandstructure}
We used DFT for the mean-field calculations as implemented in Quantum Espresso\cite{giannozzi2009quantum,giannozzi2017advanced} with the PBE functional.\cite{perdew1996generalized} We optimize the atomic positions in the unit cell with fixed lattice parameters, as defined in Table ~\ref{tab:material_parameters}. The self-consistent field calculation are performed with the following $k$-grids: $6\times6\times6$ $\Gamma$-centered for cubic MgO, $6\times6\times6$ half-shifted for zincblende CdS, $6\times6\times6$ half-shifted for cubic AgCl, and $6\times4\times6$ half-shifted for orthorhombic CsPbBr$_3$. 

With the Kohn-Sham wavefunctions from the DFT calculations, we conduct single-shot G$_0$W$_0$ calculations using BerkeleyGW\cite{deslippe2012berkeleygw} with a generalized plasmon pole model\cite{hybertsen1986electron} to compute the dielectric function at finite frequencies. We converged the quasiparticle bandgaps within 0.1 eV. For MgO, we used 600 bands and a 50 Ry polarizability cutoff. For zincblende CdS, we used 600 bands and 40 Ry. For AgCl, we used 600 bands and 50 Ry. The effective masses were obtained from finite-difference calculations near the conduction band minimum and valence band maximum,
\begin{equation} \label{eff_mass}
    m_{i} \approx  \bigg( \frac{E(\Delta \mathbf{k}) + E(-\Delta \mathbf{k}) - 2 E(0)}{\hbar^2 \Delta \mathbf{k}^2}\bigg)^{-1}
\end{equation}
\newline
\noindent
where we choose $\Delta \mathbf{k} = 0.01$ in crystal coordinates and average across the three axes.\cite{alvertis2024phonon} In cases where the band extrema are degenerate, we use the effective mass of the heaviest band, as it couples most strongly to the lattice. A more general treatment would employ the Luttinger-Kohn parameterization of degenerate band manifolds.\cite{luttinger1955motion} All of the effective masses were evaluated from single-shot GW, with the exception of CsPbBr$_3$\cite{filip2021phonon} for which we use previously reported band structures. The single particle gap, electron and hole masses are summarized in Table ~\ref{tab:material_parameters}.

\subsection{Phonon dispersion and dielectric tensor}

Phonon dispersions and dielectric tensors were computed using
density functional perturbation theory (DFPT).\cite{baroni2001phonons} Starting from the converged 
Kohn-Sham wavefunctions and charge densities obtained in the 
self-consistent field calculations described in the previous section, 
we computed the dynamical matrices on a uniform grid of 
\textbf{q}-points in the Brillouin zone. For MgO, CdS, and AgCl, a
$6\times6\times6$ \textbf{q}-point grid was employed. Due to the
larger unit cell and associated computational cost, a
$2\times2\times2$ \textbf{q}-point grid was used for orthorhombic
CsPbBr$_3$.

The dynamical matrices computed on these coarse \textbf{q}-point grids
were Fourier interpolated to obtain the interatomic force constants in
real space. For polar materials, the non-analytic correction to the 
dynamical matrix at $\mathbf{q}\to\mathbf{0}$, which 
gives rise to the LO--TO splitting at the $\Gamma$ point, 
was accounted for using the Born effective charges 
$\mathbf{Z}^*_\kappa$ and the electronic dielectric tensor 
$\boldsymbol{\varepsilon}_\infty$, both of which are 
computed self-consistently within the DFPT framework at 
$\mathbf{q}=\mathbf{0}$ via the linear response of the 
system to a homogeneous electric field 
perturbation.\cite{baroni2001phonons,gonze1997dynamical} Phonon dispersions along high-symmetry paths in the 
Brillouin zone were then obtained by inverse Fourier transformation 
of the real-space force constants. The crystalline acoustic sum rule\cite{baroni2001phonons,gonze1997dynamical} 
was imposed during the interpolation to enforce the translational 
invariance of the force constants, ensuring that the three acoustic 
branches vanish at the $\Gamma$ point as required by symmetry. This 
correction redistributes small residual violations of translational 
invariance arising from finite \textbf{q}-point sampling and numerical 
precision across all interatomic force constants in a manner 
consistent with the crystal symmetry.

In the long-wavelength limit, the acoustic phonon branches exhibit a 
linear dispersion. 
We extract $v_s$ for each acoustic branch 
by fitting the computed dispersion along each high-symmetry path in the 
vicinity of the $\Gamma$ point. Since the speed of sound generally 
depends on the propagation direction due to elastic anisotropy, we 
obtain an isotropic effective speed of sound by performing a weighted 
average over symmetry-equivalent paths.\cite{li2021deformation} For face-centered cubic 
systems, this takes the form
\begin{equation} \label{avg_vs}
    v_s = \sqrt{\frac{n_{\Gamma\text{-}X}\,v_{s,\Gamma\text{-}X}^2 
    + n_{\Gamma\text{-}L}\,v_{s,\Gamma\text{-}L}^2
    + n_{\Gamma\text{-}K}\,v_{s,\Gamma\text{-}K}^2}
    {n_{\Gamma\text{-}X} + n_{\Gamma\text{-}L}
    + n_{\Gamma\text{-}K}}}
\end{equation}
where the number of equivalent symmetry paths are 
$n_{\Gamma\text{-}X} = 6$, $n_{\Gamma\text{-}L} = 8$, and 
$n_{\Gamma\text{-}K} = 12$. For orthorhombic CsPbBr$_3$, we average 
over the $\Gamma\text{-}X$, $\Gamma\text{-}Y$, and $\Gamma\text{-}Z$ 
paths. In contrast to the acoustic branches, the optical phonon 
branches exhibit approximately flat dispersions near the zone center. 
We therefore use the zone-center ($\mathbf{q} = \mathbf{0}$) 
frequencies $\omega_{\text{LO}}$ and $\omega_{\text{TO}}$ for the 
optical modes throughout this work.

\begin{table}
\caption{\centering electron--phonon Coupling Parameters}
\label{tab:acoustic_params}
\begin{tabular}{lcccc} 
\hline
Parameter & \: MgO & \: CdS & \: AgCl & \: CsPbBr$_3$ \\  
\hline\hline
$v_{\text{LA}}$ (km/s) & \: 10.06 & \: 4.16 & \: 3.06 & \: 4.39 \\
$v_{\text{TA}_1}$ (km/s) & \: 6.56 & \: 1.60 & \: 1.12 & \: 1.86 \\
$v_{\text{TA}_2}$ (km/s) & \: 7.00 & \: 2.14 & \: 1.18 & \: 2.57 \\
$D^{\text{LA}}_e$ (eV) & \: 1.56 & \: 7.43 & \: 2.19 & \: 0.31 \\
$D^{\text{LA}}_h$ (eV) & \: 4.82 & \: 3.21 & \: 1.12 & \: 0.22 \\
$D^{\text{TA}_1}_e$ (eV) & \: 0.0 & \: 0.0 & \: 0.0 & \: 0.007 \\
$D^{\text{TA}_1}_h$ (eV) & \: 0.0 & \: 0.0 & \: 0.0 & \: 0.020 \\
$D^{\text{TA}_2}_e$ (eV) & \: 0.24 & \: 2.00 & \: 1.53 & \: 0.012 \\
$D^{\text{TA}_2}_h$ (eV) & \: 0.73 & \: 1.20 & \: 0.92 & \: 0.009 \\
$\hbar \omega_{\text{LO}_1}$ (meV) & \: 84 & \: 34 & \: 20 & \: 18 \\
$\hbar\omega_{\text{LO}_2}$ (meV) & \: -- & \: -- & \: -- & \: 11 \\
$\hbar\omega_{\text{LO}_3}$ (meV) & \: -- & \: -- & \: -- & \: 7 \\
$\hbar\omega_{\text{TO}_1}$ (meV) & \: -- & \: 26 & \: -- & \: -- \\
$\hbar\omega_{\text{TO}_2}$ (meV) & \: -- & \: 26 & \: -- & \: -- \\
$\alpha_e$ & \: 1.59 & \: 0.45 & \: 1.87 & \: 2.13 \\
$\alpha_h$ & \: 6.11 & \: 1.85 & \: 3.37 & \: 2.13 \\
$C^{\text{LO}_1}_{e/h}$ (eV$\cdot$\AA$^{-2}$) & \: 11.70 & \: 3.00 & \: 2.80 & \: 2.45 \\
$C^{\text{LO}_2}_{e/h}$ (eV$\cdot$\AA$^{-2}$) & \: -- & \: -- & \: -- & \: 0.11 \\
$C^{\text{LO}_3}_{e/h}$ (eV$\cdot$\AA$^{-2}$) & \: -- & \: -- & \: -- & \: 0.27 \\
$D^{\text{TO}_1}_e$ (eV) & \: 0.0 & \: 0.024 & \: 0.0 & \: 0.0 \\
$D^{\text{TO}_1}_h$ (eV) & \: 0.0 & \: 0.240 & \: 0.0 & \: 0.0 \\
$D^{\text{TO}_2}_e$ (eV) & \: 0.0 & \: 0.024 & \: 0.0 & \: 0.0 \\
$D^{\text{TO}_2}_h$ (eV) & \: 0.0 & \: 0.240 & \: 0.0 & \: 0.0 \\
\hline
\end{tabular}
\end{table}

In principle, the 
high-frequency dielectric tensor can also be obtained within the GW 
framework, where many-body exchange-correlation effects beyond the 
random-phase approximation used in density functional perturbation theory are accounted for. However, 
for the systems studied in this work, we find the differences between the DFPT and GW values 
of $\varepsilon_\infty$ (taken as the isotropic average of 
the three diagonal elements of 
the dielectric constant tensor) 
to be insignificant, and we therefore use the DFPT values 
throughout.

The macroscopic low-frequency dielectric tensor
$\boldsymbol{\varepsilon}_\mathrm{s}$ was obtained by combining the electronic contribution
$\boldsymbol{\varepsilon}_\infty$ with the ionic contribution, the latter
being determined from the Born effective charges and the zone-center
phonon frequencies. Specifically, the static dielectric tensor can be
written as\cite{baroni2001phonons,gonze1997dynamical}
\begin{equation}
  \varepsilon_\mathrm{s}^{\alpha,\beta} = \varepsilon_\infty^{\alpha\beta}
  + \frac{4\pi}{V} \sum_n
  \frac{S_{n,\alpha\beta}}{\omega_n^2},
\end{equation}
where $\omega_n$ are the
transverse-optical phonon frequencies at the $\Gamma$ point, and
$S_{n,\alpha\beta}$ is the mode oscillator strength tensor defined by
\begin{equation}
  S_{n,\alpha\beta} = \left(\sum_{\kappa\gamma}
  \frac{Z^*_{\kappa,\alpha\gamma}\, e_{\kappa\gamma,n}}
  {\sqrt{m_\kappa}}\right)
  \left(\sum_{\kappa'\gamma'}
  \frac{Z^*_{\kappa',\beta\gamma'}\, e_{\kappa'\gamma',n}}
  {\sqrt{m_{\kappa'}}}\right),
\end{equation}
with $e_{\kappa\gamma,n}$ denoting the $\gamma$ Cartesian component of 
the phonon eigenvector for atom $\kappa$ in mode $n$, 
$Z^*_{\kappa,\alpha\gamma}$ the $\alpha\gamma$ component of the Born 
effective charge tensor for atom $\kappa$, and $m_\kappa$ the atomic 
mass. These quantities are directly available from the DFPT 
calculation.

\subsection{Electron--phonon interactions}

Generically, the electron--phonon interactions are evaluated from density 
functional perturbation theory. In particular, the matrix elements are 
given by\cite{savrasov1994linear}
\begin{equation} \label{eph_gen}
    g^{n}_{\lambda}(\mathbf{k},\mathbf{q}) = \sqrt{\frac{\hbar}{2 \rho V \omega_{\mathbf{q},n}}
    } \langle \lambda\mathbf{k}+\mathbf{q} | 
    \Delta_{\mathbf{q}}^{n}V_{KS}(\mathbf{r})| \lambda\mathbf{k} \rangle 
\end{equation}
where $\omega_{\mathbf{q},n}$ is the phonon frequency for mode $n$, and 
$\Delta_{\mathbf{q}}^{n}V_{KS}(\mathbf{r})$ is the linear response of 
the Kohn-Sham potential. Here, $\lambda$ is the electronic band index. 
Since we consider Wannier--Mott excitons composed of a single conduction 
band electron and a single valence band hole, we restrict ourselves to 
intraband couplings evaluated at the conduction band minimum 
and valence band maximum, respectively. For plotting and fitting the 
electron--phonon couplings, it is convenient to define\cite{li2021deformation}
\begin{equation} \label{eph_m}
    M^{n}_{\lambda}(\mathbf{q}) = \sqrt{\frac{2 \rho V\omega_{\mathbf{q},n}}
    {\hbar}}\, g^{n}_{\lambda}(\mathbf{q})
\end{equation}
which isolates the matrix element. We computed $g^{n}_{\lambda}(\mathbf{q})$ for the 
valence band maximum and conduction band minimum using the EPW package.\cite{ponce2016epw} The electron--phonon matrix elements were first computed 
on the same coarse $\mathbf{q}$-point grids used for the 
phonon calculations and then Wannier--Fourier interpolated\cite{giustino2007electron} onto 
dense $\mathbf{q}$-grids along the high--symmetry paths.

For LO phonons in polar materials, the charge-phonon matrix element is 
well approximated and fitted in the long-wavelength limit by
\begin{equation} \label{LO_M}
    M^\mathrm{LO}_\lambda(\mathbf{q}) = \frac{C^{\text{LO}}_\lambda}{q}
\end{equation}
\noindent
In the case of a single isotropic LO mode coupling to a single 
parabolic band, the constant is well approximated by the Fröhlich model\cite{frohlich1954electrons}
\begin{equation} \label{LO_fro}
    C^{\text{LO}}_\lambda \approx \hbar \omega_{\text{LO}}  
    \left(\frac{\hbar}{2 m_\lambda \omega_{\text{LO}}}\right)^{1/4} 
    \left (4\pi \alpha_\lambda \right)^{1/2} 
    \left(\frac{2 \rho  \omega_{\text{LO}}}{\hbar}\right)^{1/2} 
\end{equation}
with
\begin{equation} \label{alp}
    \alpha_\lambda = \frac{e^2}{4\pi \epsilon_0 \hbar}
    \sqrt{\frac{m_\lambda}{2\hbar \omega_{\text{LO}}}} 
    \left(\frac{1}{\epsilon_\infty} - \frac{1}{\epsilon_\mathrm{s}}\right)
\end{equation}
which results from simple dielectric continuum theory. The effective non-local interaction between imaginary time slices is
\begin{equation} \label{LO_IF}
    G_{cd}^{\text{LO}} = \frac{\Gamma_{cd} C^{\text{LO}}_c C^{\text{LO}}_d}
    {16\pi \rho \omega_{\text{LO}}} \frac{\chi^\mathrm{LO}_{|\tau-\tau'|}}{|\mathbf{x}_{c,\tau}-\mathbf{x}_{d,\tau'}|}\, 
\end{equation}
a retarded Coulomb potential.


For short-range couplings such as the deformation potential, the 
charge-phonon matrix elements can vary along different symmetry paths 
in the Brillouin zone. This anisotropy arises from the directional 
dependence of the deformation potential tensor, which reflects how 
band energies shift under strain applied along different 
crystallographic directions. We obtain isotropic effective coupling 
constants by performing a weighted average over symmetry-equivalent 
paths analogous to Eq.~\ref{avg_vs}
\begin{equation} \label{avg_D}
    D_\lambda^n = \sqrt{\frac{\sum_i n_i (D_{\lambda,i}^n)^2}{\sum_i n_i}}
\end{equation}
where $n_i$ is the multiplicity of path $i$ and $D_{\lambda,i}^n$ is the 
deformation potential fitted along that path for mode $n$. 

For acoustic phonons, the charge-phonon matrix element is linear at 
lowest order and fitted by
\begin{equation} \label{Ac_M}
    M_\lambda^\mathrm{A}(\mathbf{q}) = D_\lambda^\mathrm{A}  q 
\end{equation}
in the long-wavelength limit. The effective nonlocal interaction 
resulting from this coupling is
\begin{equation} \label{deform_IF}
    G^{\text{A}}_{cd} = \frac{\Gamma_{cd}D_c^\text{A} D_d^\text{A}}{8\pi^2 \rho v_s} 
    \int_{0}^{q_\mathrm{o}} \frac{q^2 
    \sin(q |\mathbf{x}_{c,\tau}-\mathbf{x}_{d,\tau'}|)}
    {|\mathbf{x}_{c,\tau}-\mathbf{x}_{d,\tau'}|}
     \chi_{|\tau-\tau'|}^{\mathrm{A},q}\, dq
\end{equation}
which has an ultraviolet divergence at equal imaginary times away from 
$0$\,K due to the Bose occupation factor. We regularize this integral by 
introducing a cutoff $q_\mathrm{o}$ from a Debye 
sphere\cite{fantoni2012localization,fantoni2013low}
\begin{equation} \label{debye_sphere}
    q_\mathrm{o} = \bigg(\frac{6 \pi^2}{V} \bigg)^{1/3}
\end{equation}
but find our exciton binding results are not sensitive to this particular 
choice. Analogously, at lowest order the coupling to the TO mode is constant and given by
\begin{equation} \label{TO_M}
    M_\lambda^\mathrm{TO}(\mathbf{q}) = D^\text{TO}_\lambda
\end{equation}
with a resulting non-local interaction for the charge-TO phonon deformation
\begin{equation} \label{deform_TO}
    G_{cd}^{\text{TO}} = \frac{\Gamma_{cd} D_c^{\text{TO}} D_d^{\text{TO}}\chi_{|\tau-\tau'|}^{\mathrm{TO}}}
    {8\pi^2 \rho\, \omega_{\text{TO}}} \int_{0}^{q_\mathrm{o}} 
    \frac{q\, \sin(q|\mathbf{x}_{c,\tau}-\mathbf{x}_{d,\tau'}|)}
    {|\mathbf{x}_{c,\tau}-\mathbf{x}_{d,\tau'}|}
     \, dq
\end{equation}
For the systems studied across this work, only CdS has non-zero 
transverse optical phonon couplings at lowest order. This is a 
consequence of this system not being centrosymmetric.


\section{PARAMETERIZATION of The HAMILTONIANs}
We begin by presenting the electronic structure calculations and fits used to parameterize our total action in Eqn.~\ref{influence_general_compact}. The specific values of the parameters used for all four systems in the PIMC algorithm are summarized in Tables~\ref{tab:material_parameters} and \ref{tab:acoustic_params}. We note that the quasiparticle effective masses, bandgaps, and optical dielectric constants from Table~\ref{tab:material_parameters} are plugged directly into the electronic piece of the action for the PIMC algorithm, Eqn.~\ref{elec_act_dis}, while the parameters from Table~\ref{tab:acoustic_params} are used in the non-local piece of the action resulting from integrating out the phonon degrees of freedom, Eqn.~\ref{elec_act_dis}. The systems we consider span different unit cell symmetry groups and sizes, and includes direct and indirect band gap materials. All are relatively polar, and host Wannier--Mott type excitons. 

\begin{figure}[t]
\includegraphics[width=\columnwidth]{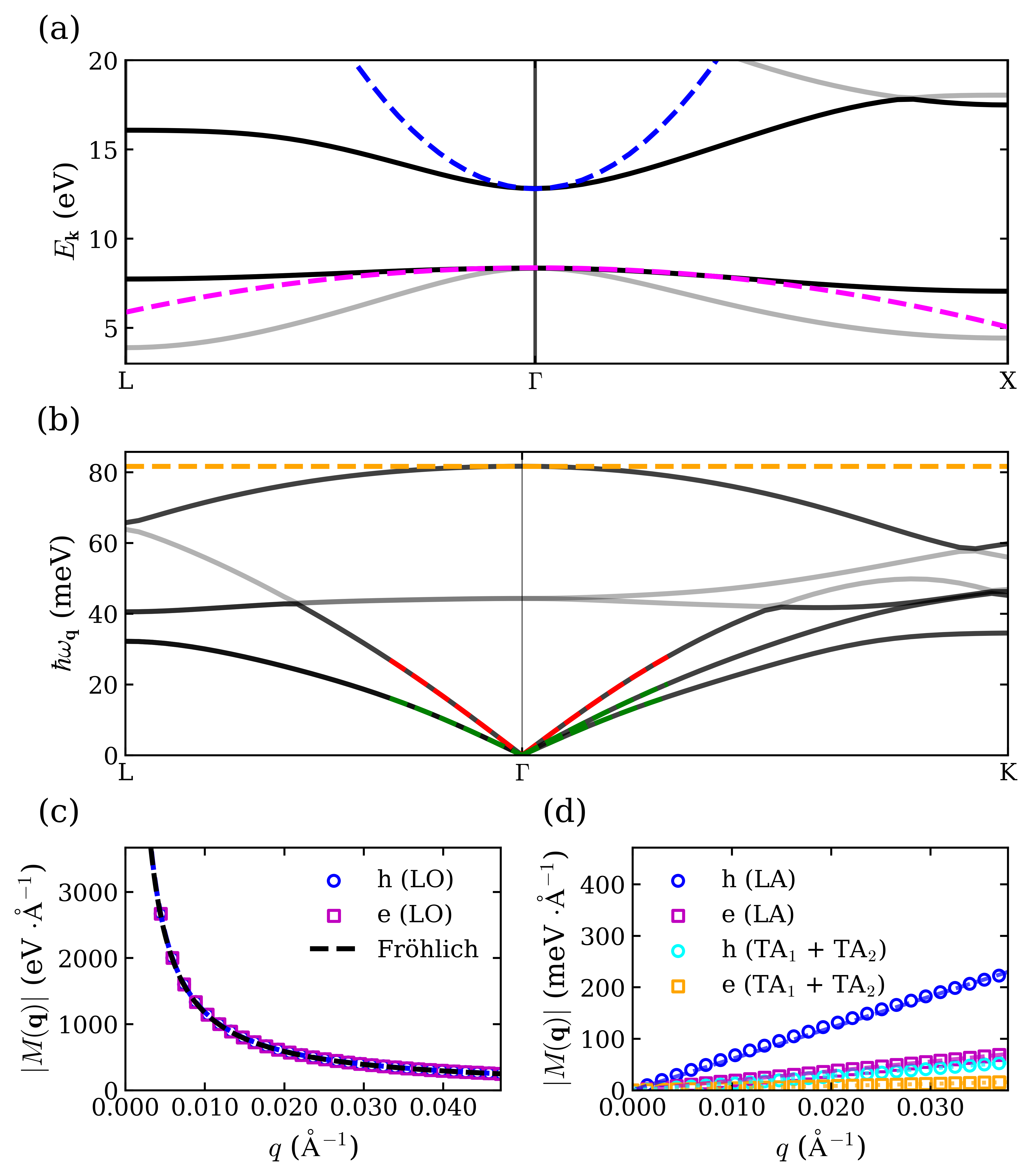}
\caption{(a) MgO electronic band structure with effective mass fits at the valence band maximum and conduction band minimum, (b) MgO phonon band structure with dashed lines illustrating the long-wavelength mode behaviors, (c) MgO LO mode interactions fits in the $\Gamma-X$ direction, and (d) MgO acoustic mode interactions fits in the $\Gamma-X$ direction.}
\label{mgo_es}
\end{figure}
\subsection{MgO}

Magnesium Oxide, MgO, crystallizes in the rock-salt structure with a face-centered cubic
lattice.\cite{slack1962thermal} The band structure, shown in Fig.~\ref{mgo_es}a, reveals a
wide direct band gap of 7.95~eV at the $\Gamma$ point. The conduction
band minimum exhibits a nearly parabolic dispersion with an effective
mass of $m_e = 0.34~m_0$. The conduction band states are primarily
derived from Mg $3s$ orbitals. The valence band maximum is
significantly flatter, yielding a much heavier hole with effective mass of
$m_h = 5.0~m_0$, and valence band states predominantly composed
of O $2p$ orbitals. This strong asymmetry between the electron and hole
masses has important consequences for polaron formation, as the heavier
hole couples more strongly to the lattice and forms a more deeply bound
polaron.

The phonon dispersion, shown in Fig.~\ref{mgo_es}b, reflects the
two-atom primitive cell, giving rise to three acoustic and three optical
branches. The large mass contrast between Mg and O, combined with the
strong ionic bonding, produces a wide phonon band gap between the
acoustic and optical branches. The LO mode at the $\Gamma$ point has an
energy of 84~meV, which is notably higher than in the other systems studied in
this work. The large LO--TO splitting, driven by the significant Born
effective charges and the relatively low high-frequency dielectric
constant, $\varepsilon_\infty = 3.3$, reflects the strongly polar
character of MgO. The acoustic branches exhibit stiff dispersions, with
averaged speeds of sound of $v_{\text{LA}} = 10.06$ Km/s,
$v_{\text{TA}_1} = 6.56$ Km/s, and $v_{\text{TA}_2} = 7.00$ Km/s,
which are the highest among the four systems studied and reflect the
light atomic masses and strong bonding in MgO.

Figures~\ref{mgo_es}c and \ref{mgo_es}d show that MgO exhibits
significant electron--phonon coupling to both the LO and acoustic modes,
which is reinforced by the large magnitudes of the fitted coupling
strengths presented in Table~\ref{tab:acoustic_params}. The coupling is much stronger to the LO mode than to the acoustic modes. In all cases, the $q$ dependence is well approximated by their limiting low $q$ forms. The resultant Fröhlich 
coupling constants are $\alpha_e = 1.59$ and $\alpha_h = 6.11$,
placing the hole firmly in the strong coupling regime. The constant coupling to the TO mode is ignored because it is negligibly small. 

With this parameterization, we find a  
zero-temperature exciton binding energy of $E_\mathrm{B}=$178~meV. This is substantially lower than that predicted from a Wannier model, $E^\mathrm{W}_\mathrm{B}=\mu e^2/16 \pi \varepsilon_0^2 \varepsilon_\infty^2 \hbar^2=382$ meV, without lattice fluctuations, thus implying that the inclusion of a dynamic lattice significantly renormalizes the zero-temperature exciton binding energy decreasing it by over 200 meV. The experimental uncertainty in the binding energy is large, with reported values between 80~meV\cite{walker1968phonon} and
$145 \pm 20$~meV.\cite{whited1973exciton} Nonetheless, our result of 178~meV is in much closer agreement with this experimental range than the static Wannier estimate. The renormalization of the exciton binding energy from the Wannier model value arises from two competing phonon-mediated effects. Polaronic effects stabilize each carrier, with polaron binding energies of $E_\mathrm{p}=$ 133 meV and 635 meV for the electron and hole, which act to increase the exciton binding. However, the cross-terms introduce a repulsive phonon-mediated screening of the electron--hole Coulomb interaction that more than compensates, leading to the net reduction in $E_\mathrm{B}$. The electron polaron $E_\mathrm{p}$ is well described by only considering the coupling to the LO mode. The hole polaron binding is a mixture of coupling to the LO mode and the acoustic modes, with a Fröhlich model accounting for 90\% of the binding energy.

\begin{figure}[b]
\includegraphics[width=\columnwidth]{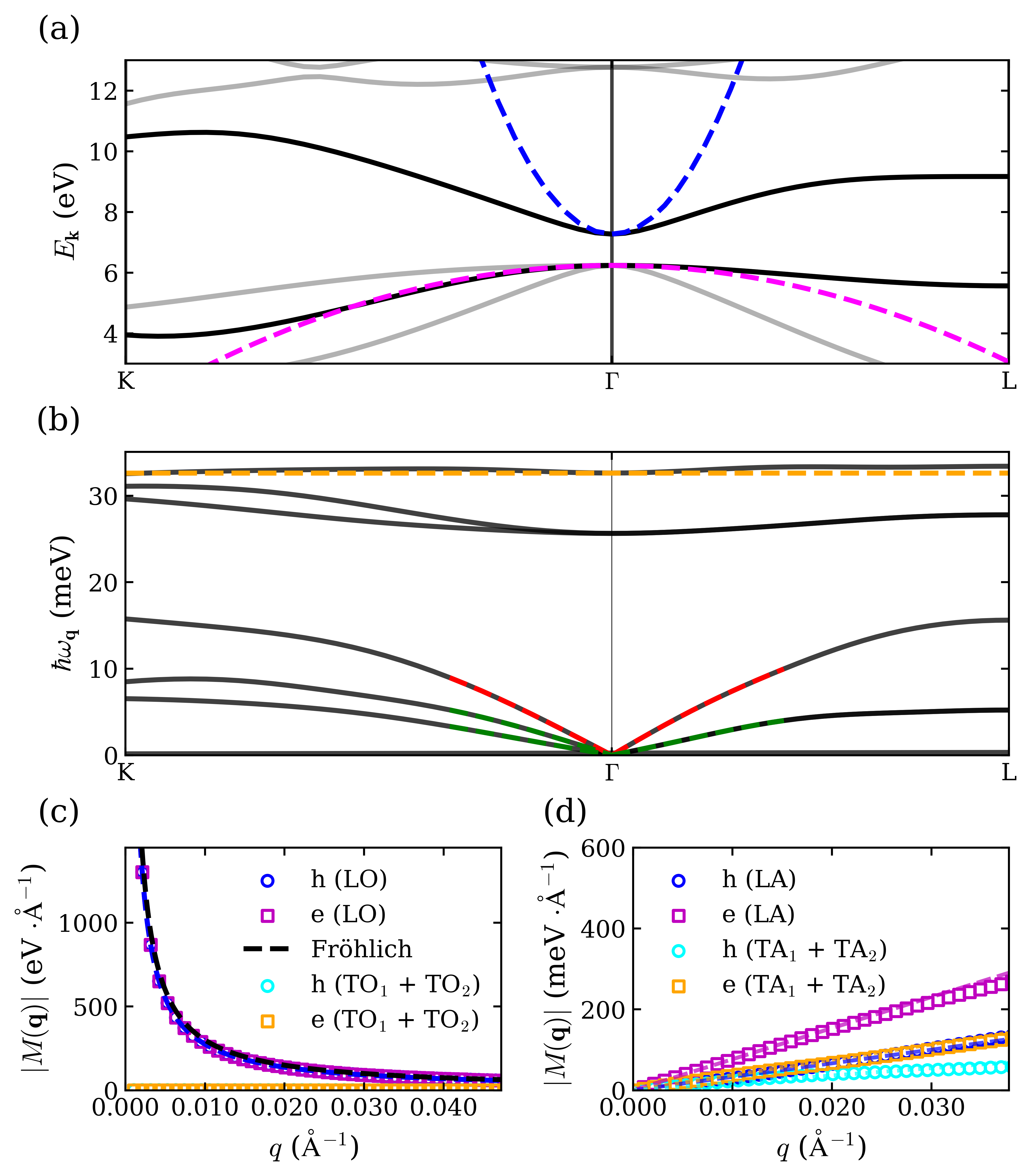}
\caption{(a) CdS electronic band structure with effective mass fits at the valence band maximum and conduction band minimum, (b) CdS phonon band structure with dashed lines illustrating the long-wavelength mode behaviors, (c) CdS optical mode interactions fits in the $\Gamma-X$ direction, and (d) CdS acoustic mode interactions fits in the $\Gamma-X$ direction.}
\label{cds_es}
\end{figure}

\subsection{CdS}
We consider Cadmium Sulfide, CdS, in the zinc-blende structure, a metastable cubic phase of this prototypical II-VI semiconductor that has attracted interest for the design of nanomaterials such as quantum dots and nanowires.\cite{murray1993synthesis,duan2000general,brus1984electron} For this reason, it has also become a model
system for computing exciton binding energies.\cite{brus1984electron,park2022renormalization,filip2021phonon,alvertis2024phonon} 
The band structure, shown in Fig.~\ref{cds_es}a, reveals a direct band
gap of 2.50~eV at the $\Gamma$ point. The conduction band minimum,
derived primarily from Cd $5s$ orbitals, is fairly non-parabolic, with
an effective mass of $m_e = 0.12~m_0$. The valence band maximum is
composed predominantly of S $3p$ orbitals and is significantly heavier,
with $m_h = 2.01~m_0$. Nonetheless, since CdS is a direct band gap
material hosting a Wannier--Mott exciton, the excitonic wavefunction is
heavily peaked at the $\Gamma$ point in momentum space, and the
parabolic effective mass approximation remains reasonable.\cite{filip2021phonon,alvertis2024phonon}

The phonon dispersion, shown in Fig.~\ref{cds_es}b, again reflects
the two-atom primitive cell, giving rise to three acoustic and three
optical branches. The large mass ratio between Cd and S produces a
clear separation between the acoustic and optical manifolds. The LO
mode at the $\Gamma$ point has an energy of 34~meV, with a moderate
LO--TO splitting reflecting the polar character of the Cd--S bond.
Although CdS exhibits two degenerate TO modes at 26~meV with non-zero
quadrupole deformation potentials, the coupling to these modes is quite
small in magnitude near the center of the Brillouin zone, as discussed
below. The acoustic branches are considerably softer than those of MgO,
with averaged speeds of sound of $v_{\text{LA}} = 4.16$ Km/s,
$v_{\text{TA}_1} = 1.60$ Km/s, and $v_{\text{TA}_2} = 2.14$ Km/s,
consistent with the heavier atomic masses and weaker bonding compared
to MgO.

Figures~\ref{cds_es}c and \ref{cds_es}d show that the charge-phonon
coupling is significant for both the LO and acoustic modes. However,
despite the non-zero quadrupolar deformation potentials for the TO
modes, the coupling to these modes remains small near the zone center.
The Fröhlich  coupling constants are $\alpha_e = 0.45$ and
$\alpha_h = 1.85$, indicating that the electron is in the weak coupling
regime while the hole is in the intermediate coupling regime. Again, we find that the $q$ dependence of all of the electron--phonon coupling elements are well described by the expected long wavelength forms. 

At 25~meV, our zero-temperature exciton binding energy for CdS is in good agreement with the experimental value of 28~meV.\cite{jakobson1994optical} 
This is about half as large
as predicted from a Wannier model without lattice fluctuations, $E_\mathrm{B}^\mathrm{W}=40$ meV. As with MgO,  CdS also exhibits an exciton binding energy renormalization, though it is a much smaller effect. As before, the dominant contribution comes from the LO mode.
The polaron binding energies are relatively small at $E_p=$ 15 meV and
68 meV for the electron and hole. The overall smaller scale is due to the smaller difference in optical and ionic dielectric constants and the lower LO frequency for CdS compared to MgO. The larger $E_\mathrm{p}$ for the hole is a consequence of its larger mass. 
Both single-carrier polaron binding energies are well
described by the Fröhlich model within first order perturbation theory $E_\mathrm{p} \approx \alpha_{i} \hbar \omega_{\mathrm{LO}}$ with predicted binding
energies of 15 meV and 63 meV.

\begin{figure}[t]
\includegraphics[width=\columnwidth]{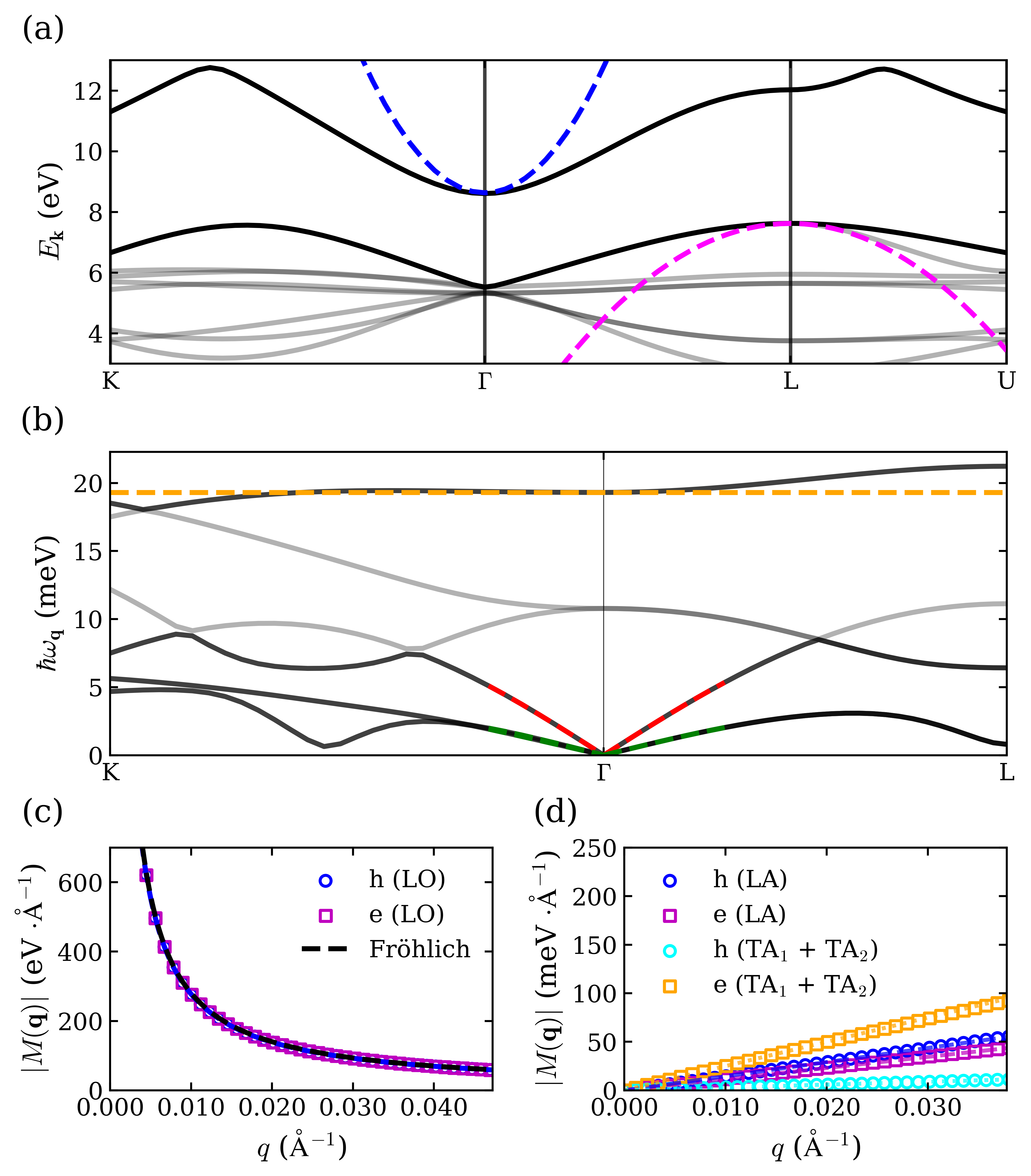}
\caption{(a) AgCl electronic band structure with effective mass fits at the valence band maximum and conduction band minimum, (b) AgCl phonon band structure with dashed lines illustrating the long-wavelength mode behaviors, (c) AgCl optical mode interactions fits in the $\Gamma-X$ direction, and (d) AgCl acoustic mode interactions fits in the $\Gamma-X$ direction.}
\label{agcl_es}
\end{figure}

\subsection{AgCl}

Silver chloride (AgCl) also crystallizes in the rock-salt structure with a
face-centered cubic lattice.\cite{aniya2012analysis} As seen from Fig.~\ref{agcl_es}a, AgCl is
an indirect bandgap material with a gap of 2.75~eV, in contrast to the
direct gap semiconductors MgO and CdS. The
valence band maximum is composed of strongly hybridized Ag $4d$ and Cl
$3p$ states, while the conduction band minimum is primarily derived
from Ag $5s$ orbitals. The conduction band minimum is
characterized by an effective mass of $m_e = 0.26~m_0$, while the
valence band maximum yields a hole effective mass of $m_h =
0.85~m_0$.

The phonon dispersion, shown in Fig.~\ref{agcl_es}b, again reflects
the two-atom primitive cell. The LO mode at the $\Gamma$ point has an
energy of 20~meV, which is the lowest among the rock-salt systems
studied here. The acoustic branches are softer than those of MgO, with
averaged speeds of sound of $v_{\text{LA}} = 3.06$ Km/s,
$v_{\text{TA}_1} = 1.12$ Km/s, and $v_{\text{TA}_2} = 1.18$ Km/s,
consistent with the heavier atomic masses. Figures~\ref{agcl_es}c and
\ref{agcl_es}d show that AgCl has significant charge-phonon coupling to
the LO mode and sizable acoustic deformation potentials, as reinforced
by the values in Table~\ref{tab:acoustic_params}. The Fröhlich 
coupling constants are $\alpha_e = 1.87$ and $\alpha_h = 3.37$,
placing both carriers in the intermediate coupling regime.

Our model predicts a zero-temperature
exciton binding energy of $E_\mathrm{B}=$60~meV. Experimental estimates of the
 indirect exciton binding energy are approximately
40~meV.\cite{von1990localized} The Wannier model without is $E_\mathrm{B}^\mathrm{W}=$108 meV, illustrating again that lattice degrees of freedom quantitatively change the exciton binding strength by over 40 meV.  For AgCl this is a significant effect, intermediate to MgO and CdS. The polaron binding energies are relatively large at $E_\mathrm{p}=$ 38 meV and 70 meV for the electron and hole. Each are completely determined by the Fröhlich coupling, and both are adequately predicted by first order perturbation theory\cite{frohlich1954electrons,feynman1955slow} yielding 37 meV and 67 meV for the electron and hole.

\begin{figure}[t]
\includegraphics[width=\columnwidth]{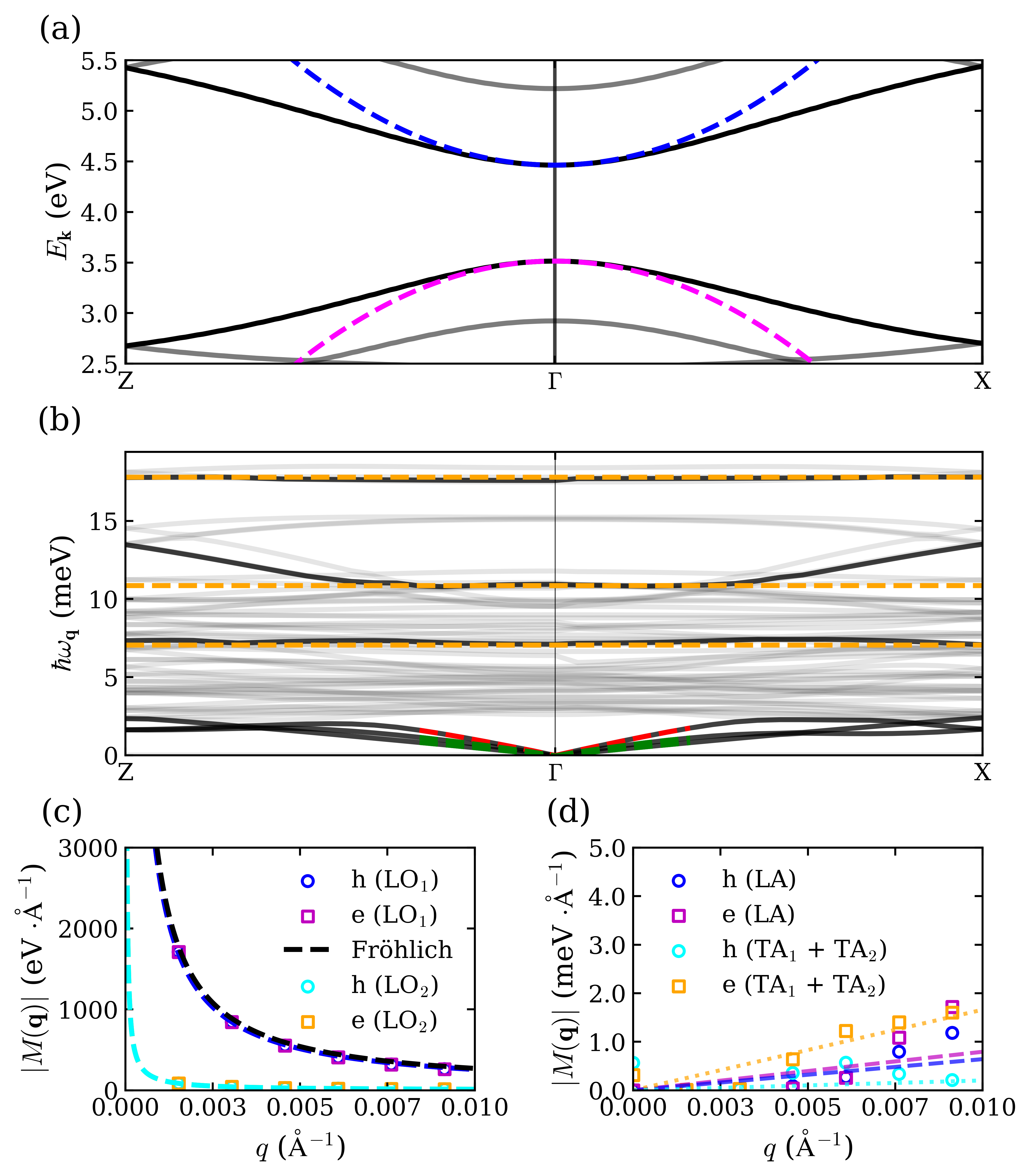}
\caption{(a) CsPbBr$_3$ electronic band structure with effective mass fits at the valence band maximum and conduction band minimum, (b) CsPbBr$_3$ phonon band structure with dashed lines illustrating the long-wavelength mode behaviors, (c) CsPbBr$_3$ LO mode interactions fits in the $\Gamma-X$ direction, and (d) CsPbBr$_3$ acoustic mode interactions fits in the $\Gamma-X$ direction. In panel (c), we note that we have not plotted the charge-phonon matrix elements for mode LO$_3$, but we have included them in the PIMC calculations.}
\label{perov_es}
\end{figure}

\subsection{CsPbBr$_3$}
We move beyond two-atom unit cells and consider orthorhombic
CsPbBr$_3$, a model halide perovskite semiconductor that has shown
great promise for applications such as solar cell design.\cite{kojima2009organometal,stoumpos2013crystal,sutton2016bandgap} CsPbBr$_3$
crystallizes in the orthorhombic $Pnma$ space group at low temperatures.\cite{stoumpos2013crystal} The considerably larger unit cell volume of 796.20~\AA$^3$
stands in contrast to the simple rock-salt and zinc-blende systems
considered above.

The band structure, shown in Fig.~\ref{perov_es}a, reveals a direct
band gap of 1.85~eV at the $\Gamma$ point. The conduction band minimum
is derived primarily from Pb $6p$ orbitals, while the valence band
maximum is composed predominantly of Br $4p$ orbitals with significant
Pb $6s$ admixture. A notable feature of CsPbBr$_3$ is the symmetry
between the electron and hole effective masses, with $m_e = m_h =
0.20~m_0$. This stands in contrast to the strong electron--hole mass
asymmetries observed in MgO, CdS, and AgCl, and reflects the
antibonding character of both band edges in lead halide perovskites.

The phonon dispersion, shown in Fig.~\ref{perov_es}b, reflects the
larger primitive cell, giving rise to a dense manifold of phonon
branches. The low-frequency optical modes characteristic of halide
perovskites reflect the softness of the lattice, which arises from the
heavy atomic masses and weak bonding in these materials. For simplicity,
we consider only the three most dominant LO modes, with energies of
$\hbar\omega_{\text{LO}_1} = 18$~meV,
$\hbar\omega_{\text{LO}_2} = 11$~meV, and
$\hbar\omega_{\text{LO}_3} = 7$~meV, along with the three acoustic
modes. These LO mode energies are significantly lower than those of MgO
and CdS, further highlighting the softness of the perovskite lattice.
The acoustic branches are characterized by speeds of sound of
$v_{\text{LA}} = 4.39$ Km/s, $v_{\text{TA}_1} = 1.86$ Km/s, and
$v_{\text{TA}_2} = 2.57$ Km/s.

Figure~\ref{perov_es}c shows that among the LO modes, the charge-phonon
coupling is dominated by a single mode with coupling strength
$C_{\text{LO}_1} = 2.45$~eV$\cdot$\AA$^{-2}$ for both the electron and hole, while the
remaining two modes contribute coupling strengths of only 0.11 and
0.27~eV$\cdot$\AA$^{-2}$, respectively. This dominance of a single LO
mode is consistent with experiment.\cite{iaru2021frohlich} It also highlights an important
limitation of the standard Fröhlich  model when applied to systems
with multiple LO phonon modes.\cite{martin2023multiple,verdi2015frohlich,frost2017calculating} The standard approach becomes inadequate
because it does not perform a proper mode-by-mode decomposition of the
lattice polarizability contributions to the static dielectric constant. There are correspondingly three distinct Fröhlich coupling constants that are the same for the electron and hole, which are $\alpha_{e,1}$ = 1.96, $\alpha_{e,2}$ =0.013, and $\alpha_{e,3}$ =0.24, suggesting that the single lowest optical frequency contributes most. 
Additionally, Fig.~\ref{perov_es}d shows that the charge-phonon
coupling to the acoustic modes is comparatively minor for CsPbBr$_3$,
with deformation potentials an order of magnitude smaller than those of
the other systems studied, shown in Table~\ref{tab:acoustic_params}. This is in
stark contrast to MgO, CdS, and AgCl, where acoustic coupling plays a
significant role. As with the other systems, the functional forms of the coupling matrix elements are well described by their limiting low $q$ forms.

The zero-temperature exciton binding energy from our dynamical lattice calculation yields a value of 40~meV, which is in good agreement with the experimental value of
33~meV\cite{yang2017impact} despite the neglect of lattice anharmonicity.\cite{park2022nonlocal} The Wannier model predicts 69 meV, demonstrating that coupling to polar phonons attenuates exciton binding, consistent with previous work.\cite{park2022renormalization} The polaron binding energies are both 38 meV, which is completely determined by coupling to optical modes, and is well approximated by weak coupling perturbation theory. 

\begin{figure}
\includegraphics[width = \columnwidth]{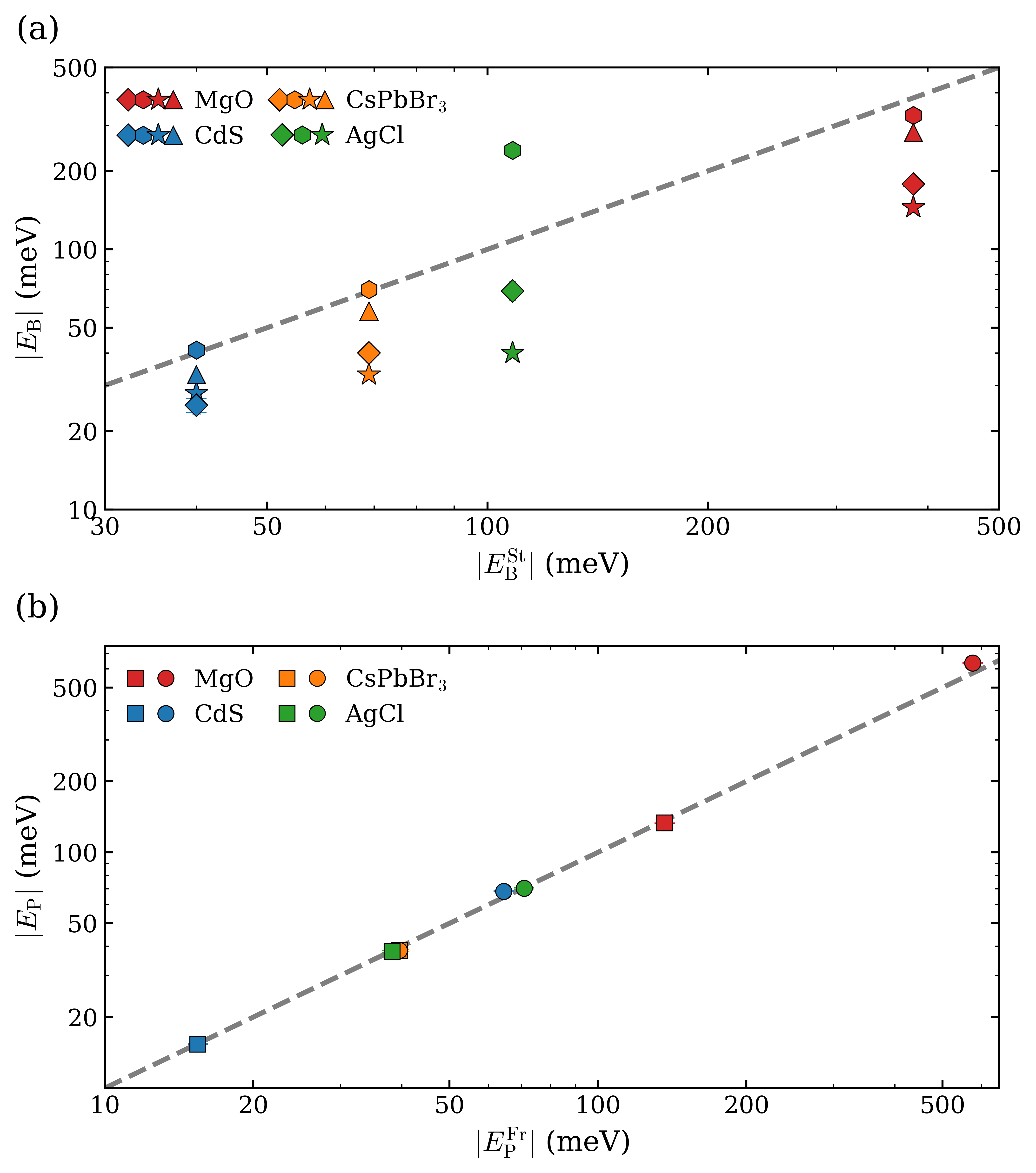}
\caption{(a) 0K Renormalized exciton binding energies versus static lattice exciton binding energies and (b) 0K Full polaron binding energies versus Fröhlich model polaron binding energies all plotted on a log-log scale. The diamond points correspond to PIMC results from this work. The starred, hexagonal, and triangular points in (a) correspond to experimental values, computed GW--BSE values, and perturbative phonon-renormalized values\cite{alvertis2024phonon,filip2021phonon}, respectively. The dashed line corresponds to $y = x$. The squares in (b) correspond to electron polarons, and the circles correspond to hole polarons.}
\label{Eb_pol_0K}
\end{figure}

\section{Quasiparticle binding}
With the Hamiltonians parameterized, we now summarize our calculations of the exciton and polaron binding energies at zero and elevated temperature. In the following, we compare our results to other models, both ab initio and those informed from continuum theories. For the large excitons considered in this work, which predominantly couple to polar phonons, the continuum theories are found to be especially appropriate.

\subsection{Zero temperature energies}
In Fig. \ref{Eb_pol_0K}a) we compare our calculated $E_\mathrm{B}$ with predictions from the simple Wannier--Mott model, as well as experimental measurements and previous calculations using GW--BSE without inclusion of phonon screening contributions.\cite{filip2021phonon,alvertis2024phonon,lorin2021first} The GW--BSE calculations largely reproduce the Wannier--Mott model.  This observation serves to calibrate our usage of an effective mass model, as our static Wannier exciton binding energies are in good agreement with GW--BSE calculations conducted with frequency-dependent dielectric tensors and full band dispersion. The sole exception is for AgCl. This discrepancy is not unexpected, as AgCl is known to be particularly challenging due to the strong hybridization of Ag $4d$ states with Cl $3p$ orbitals, which has a direct impact on the band structure and estimates of excitonic effects.\cite{lorin2021first} 

Our calculated binding energies are systematically lower than the Wannier--Mott model, confirming the intuitive expectation that including electron phonon coupling reduces the exciton binding energy. They are systematically closer to experimentally reported values than GW--BSE. Quantitatively, our calculated $E_\mathrm{B}$ agrees with the experimental values to within 11\% for CdS, 21\% for CsPbBr$_3$, 50\% for AgCl, and 23\% for MgO, where for the latter we compare to the upper bound of the reported experimental range. We have also compared to perturbative phonon-renormalized exciton binding energies\cite{alvertis2024phonon,filip2021phonon}, which include the leading-order dynamic screening to the electron--hole interaction but exclude single-carrier polaron effects. For the three materials common between the works, the perturbative values systematically overestimate $E_\mathrm{B}$ relative to both experiment and our PIMC results, yielding 33, 58, and 281~meV for CdS, CsPbBr$_3$, and MgO, respectively. The differences arise from PIMC capturing the non-perturbative resummation of the phonon-mediated one-body and two-body interactions.

In Fig. \ref{Eb_pol_0K}b) we compare our calculated electron and hole polaron binding energies, $E_\mathrm{p}$, with those computed from a simple single optical mode Fröhlich model, $E_\mathrm{p}^{\mathrm{Fr}}$. While there is no simple closed form solution for $E_\mathrm{p}^{\mathrm{Fr}}$, Feynman's two-parameter variational bound on the polaron binding energy is known to be highly accurate and is adopted here.\cite{feynman1955slow,schultz1959slow,martin2023multiple} When comparing the Fröhlich model  results at $0$ K with the PIMC results, we find nearly quantitative agreement across the range of materials studied. Slight differences arise for two main reasons. First, the effects of acoustic phonon couplings could play a role in single-carrier polaron formation, as in MgO which exhibits the largest deviation from the Fröhlich model. Second, for systems with multiple LO modes, like CsPbBr$_3$, the simple Fröhlich model breaks down, and one must properly disentangle the lattice polarization contributions to the static dielectric constant.\cite{verdi2015frohlich,frost2017calculating,martin2023multiple}


\subsection{Temperature Dependence of $E_\mathrm{B}$}
\begin{figure}
\includegraphics[width = \columnwidth]{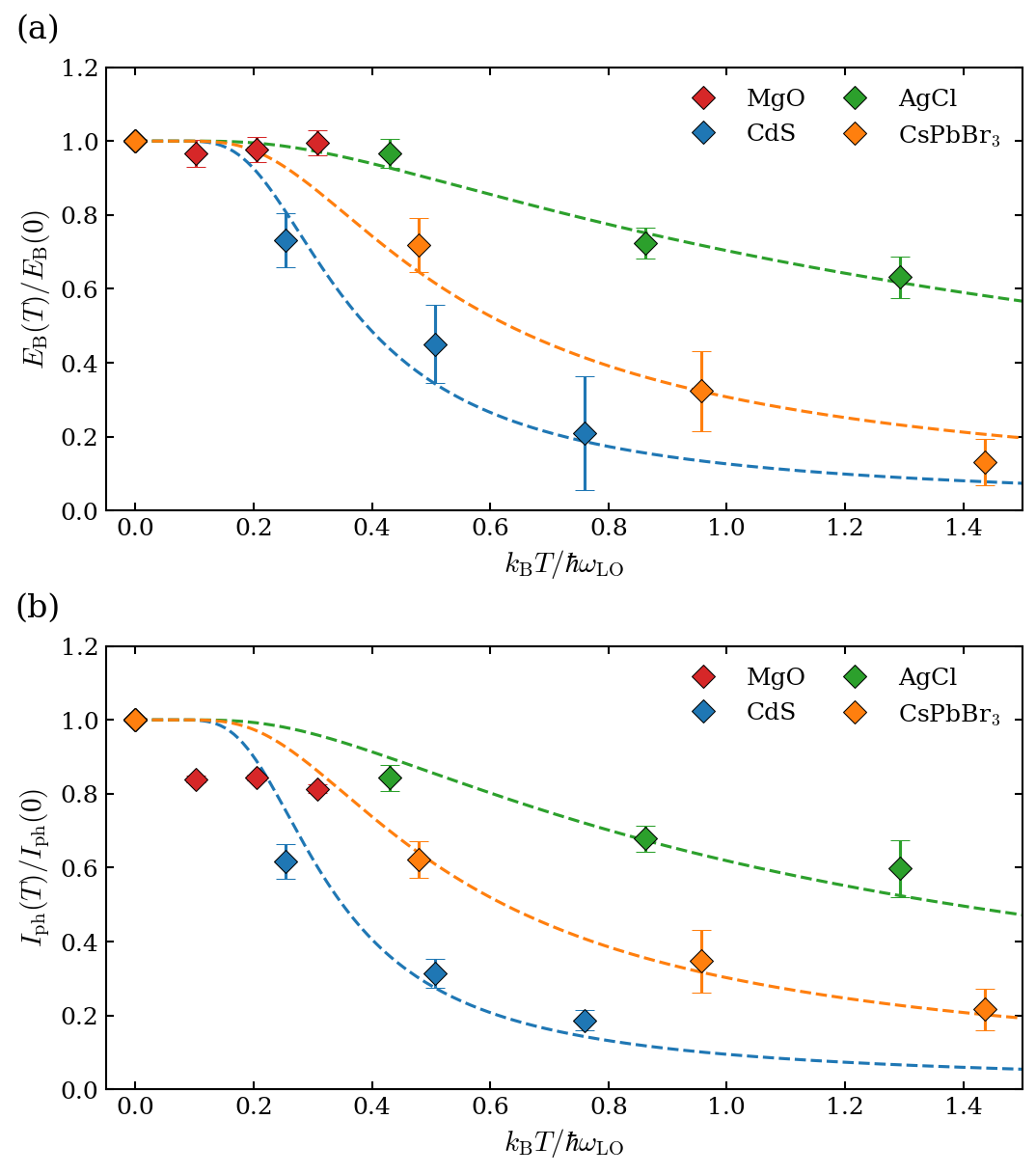}
\caption{(a) Ratio of finite temperature to 0K dynamic lattice exciton binding energy as a function of ratio of thermal energy to dominant LO phonon energy and (b) Ratio of finite temperature to 0K repulsive Coulomb screening from phonons (Eqn.~\ref{influence_general_compact}) as a function of ratio of thermal energy to dominant LO phonon energy. Solid lines represent fits to the phenomenological model (Eqn.~\ref{screening})}
\label{Eb_int_FT}
\end{figure}

PIMC allows us to straightforwardly extend our calculations of exciton binding to finite temperature. Figure~\ref{Eb_int_FT} highlights the temperature dependence from 0--300 K of the renormalized exciton binding energy across the materials studied. For CdS, AgCl, and CsPbBr$_3$, $E_\mathrm{B}$ systematically decreases with temperature. The decrease is strongest for CdS, which has the smallest $E_\mathrm{B}$ at 0 K, and the decrease is the slowest for AgCl, which of the three has the largest $E_\mathrm{B}$ at 0 K. By stark contrast, 
the MgO exciton binding energy exhibits a remarkable stability at elevated temperatures. The exciton binding energy is nearly temperature-independent in this range because it is dominated by strong static Coulomb interaction, while additional finite temperature phonon-mediated corrections remain small due to the negligible thermal occupation of the high-energy LO modes at these temperatures.

\begin{figure}
\includegraphics[width = \columnwidth]{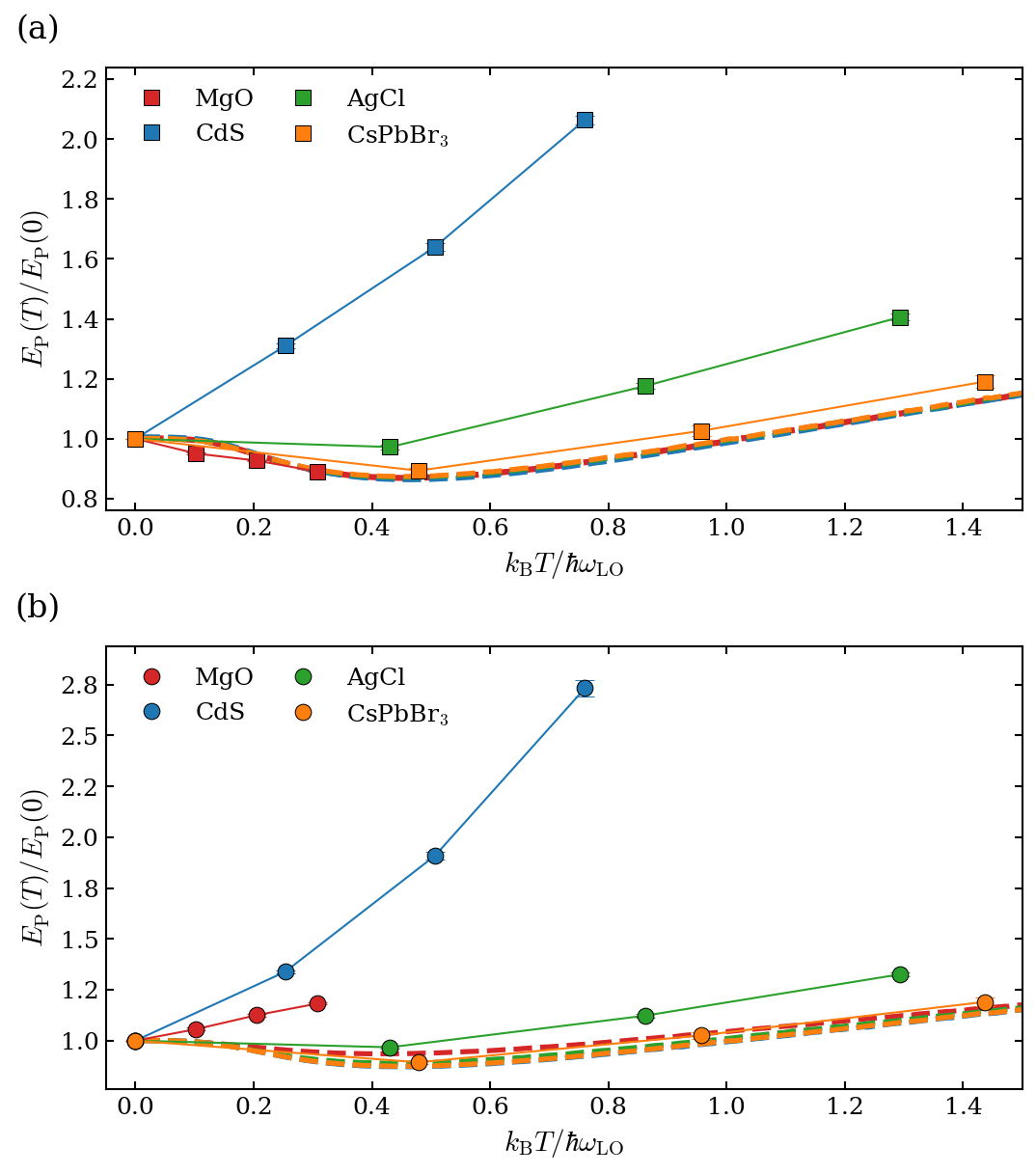}
\caption{(a) Ratio of the finite-temperature to zero-temperature full polaron binding energy as a function of ratio of thermal energy to dominant LO phonon energy for the electron and (b) for the hole. Dashed lines represent the results obtained from Feynman's variational theory.}
\label{full_pol}
\end{figure}

This temperature dependence can be understood by evaluating the temperature-dependent electron--hole interaction mediated by the exciton--phonon coupling, $I_
\mathrm{ph}$. Specifically, the effective interaction energy is given by
\begin{equation}
I_{\mathrm{ph}} = \left\langle \frac{\partial (\beta \mathcal{H}_{\text{nl}}^{c \neq d})}{\partial \beta} \right\rangle
\end{equation}
where $\mathcal{H}_{\text{nl}}^{c \neq d} = \frac{1}{\beta\hbar}\mathcal{S}_{\text{nl}}^{c \neq d}$ is the cross-term ($c \neq d$) contribution to the influence functional in Eq.~\ref{influence_general_compact}, and the expectation value is taken over the full ensemble defined by Eqn.~\ref{Partition}.
This quantity measures the repulsive electron--hole interaction screened by the phonons. The single-carrier polaron binding energies are also temperature-dependent, as shown in Fig.~\ref{full_pol}. However, because the binding energy is defined as the difference between the free two-polaron state and the bound exciton-polaron state, the self-energy contributions largely cancel. Moreover, the single-carrier polaron binding energies are only weakly temperature-dependent in the range considered here, so the residual temperature dependence of $E_\mathrm{B}$ is dominated by the cross-term $I_\mathrm{ph}$. The temperature dependence of $I_\mathrm{ph}$ mirrors almost exactly the temperature dependence of $E_\mathrm{B}$. For  CdS, AgCl, and CsPbBr$_3$ these materials show a more gradual temperature dependence where the repulsive phonon screening decreases in magnitude smoothly as
temperature increases. 
The observed temperature-dependent behavior is well reproduced with a
phenomenological thermal screening model. In particular, we define $E_\mathrm{B}(T)=E_\mathrm{B}(0)+\Delta E_\mathrm{B}(T)$, and note that from Fig.~\ref{Eb_int_FT}, $\Delta E_\mathrm{B}(T) \propto I_\mathrm{ph}(T)$. Further, we note that for all of the systems studied the dominant phonon contribution is from a single LO mode. Noting that the phonon correlator for a dispersionless optical mode is independent of $\mathbf{q}$ and can be decomposed exactly as
\begin{equation}
\chi^\mathrm{LO}_{|\tau-\tau'|} = e^{-\omega_{\mathrm{LO}}|\tau-\tau'|} + 2 n_{\mathrm{BE}} \cosh(\omega_{\mathrm{LO}}|\tau-\tau'|) ,
\end{equation}
where $n_{\mathrm{BE}} = 1/(\exp[\beta\hbar\omega_{\mathrm{LO}}] - 1)$ is the Bose occupation factor. For  coupling to the optical mode, $g^\mathrm{LO}_\lambda(\mathbf{q}) \propto 1/q$, so the momentum integral in $I_\mathrm{ph}$ can be done, yielding an expression of the temperature-dependent phonon correction to the binding energy 
\begin{equation}
\begin{split}
\Delta E_\mathrm{B}(T) = -n_{\mathrm{BE}}(T) \frac{C_e^\mathrm{LO}C_h^\mathrm{LO}}{8 \pi \rho \omega_\mathrm{LO}} W(T)
\end{split}
\end{equation}
with 
$$
W(T)=\int_\tau \int_{\tau'} \, \cosh(\omega_{\mathrm{LO}}|\tau-\tau'|)  \left\langle \frac{1}{|\mathbf{x}_{e,\tau} - \mathbf{x}_{h,\tau'}|} \right\rangle_{eh}
$$
where the temperature dependence enters through $n_{\mathrm{BE}}$, the integral $W(T)$ via its limits of integration, and the averaging operation. For a Coulomb bound state, all spatial expectation values $\langle 1/r \rangle$ scale $\propto E_\mathrm{B}(T)$. This scaling holds at each value of $|\tau-\tau'|$ independently, so the full double integral inherits it
\begin{equation}
\Delta E_\mathrm{B}(T) = -A \, n_{\mathrm{BE}} \, E_\mathrm{B}(T),
\end{equation}
where $A$ absorbs the $\tau$-dependent structure of the integral and other constants, and we ignore the small temperature dependence in the limits of integration. The binding energy then satisfies
\begin{equation}
E_\mathrm{B}(T) =  E_\mathrm{B}(0) - A \, n_{\mathrm{BE}} \, E_\mathrm{B}(T),
\end{equation}
which gives
\begin{equation}\label{screening}
\frac{E_\mathrm{B}(T)}{E_\mathrm{B}(0)} = \frac{1}{1 + A \, n_{\mathrm{BE}}}.
\end{equation}
which is plotted in Fig.~\ref{Eb_int_FT}.  The fitted values of $A$ are
consistent between $E_B$ and $I_{ph}$ within each material, suggesting
that the temperature dependence of the binding energy is closely linked
to the thermal screening of the repulsive phonon-mediated interaction.
The magnitude of $A$ reflects the relative importance of phonon-mediated
screening compared to the static Coulomb binding. Materials where the
Coulomb binding is weak relative to the phonon screening, such as CdS and CsPbBr$_3$,
exhibit larger $A$, while materials with stronger Coulomb binding, such
as AgCl, exhibit smaller $A$. This has direct consequences for the thermal equilibrium between excitons and free carriers. In the low excitation regime, the equilibrium concentration of free carriers is proportional to $\rho_\mathrm{ex}^{1/2} \exp[\beta E_\mathrm{B}(T)/2]$ where $\rho_\mathrm{ex}^{1/2}$ is the number of excitons.\cite{gourley1982experimental,chemla2003room} Thus, the dramatic decrease in binding strength with temperature of many of these materials signals the spontaneous generation of free carriers.

\subsection{Factors Influencing Polaron Formation}

To understand the temperature dependence of the polaron binding, we compare our calculated $E_\mathrm{p}$ to Feynman's variational bound of the Fröhlich model.\cite{feynman1955slow,schultz1959slow,martin2023multiple,frost2017calculating} Since the variational estimate is known to be accurate for polarons described by a Fröhlich model,\cite{hahn2018diagrammatic} deviations from it in our calculations highlight the role of acoustic phonons on polaron formation. 

For MgO, the charge-phonon coupling to the LO mode is the main mechanism for polaron formation for the electron, whereas the charge-phonon coupling to the acoustic modes becomes more dominant for the hole as temperature increases. This explains why the polaron binding energy increases in magnitude for the hole but decreases for the electron across the 0 to 300 K temperature range. For CdS, both of the quasiparticle polaron binding energies become dominated by coupling to the acoustic modes as the temperature increases, deviating markedly from expectations from the Fröhlich model.

AgCl and CsPbBr$_3$ both exhibit non-monotonic trends in polaron binding energy as a function of temperature, following the expectation from the Fröhlich model. The LO phonon energy at the $\Gamma$ point in AgCl is 20 meV, so its thermal occupation remains negligible below roughly 230 K ($k_\mathrm{B}T/\hbar\omega_{\mathrm{LO}} \approx 1$). Therefore, if only the LO mode  were considered in the polaron formation of this system, the magnitude of the polaron binding energy would be expected to decrease until around 230 K ($k_\mathrm{B}T/\hbar\omega_{\mathrm{LO}} \approx 1$) due to a weaking attraction between the bare charge carrier and the lattice distortion as the charge becomes thermally delocalized. The increase in the population of the acoustic modes overcomes this weakened attraction to the LO phonon polarization leading to an increase in the polaron binding energy starting after 100 K ($k_BT/\hbar\omega_{\mathrm{LO}} \approx 0.4$). For CsPbBr$_3$, the dominant LO mode starts to become populated around 200 K ($k_BT/\hbar\omega_{\mathrm{LO}} \approx 1$), which explains why the polaron binding energy increases in magnitude starting around that temperature range. Deviations from the single LO-mode Fröhlich model results from the two weaker LO modes becoming populated. 

In addition to the competition between the carrier coupling to the LO and acoustic modes, there is also a competition between the population of the phonon modes and the bare interaction strength of a charge carrier with the lattice as a function of temperature. This behavior is also  captured by Feynman's variational bound. However, the variational bound approximates the retarded self-interaction of the carrier as quadratic in imaginary time, and this approximation is least accurate at low temperatures in the frozen phonon regime, where the carrier is thermally sampling excited polaron configurations but the phonon field remains in its ground state. In this regime, the true self-interaction, which is non-quadratic in the carrier coordinates, is poorly approximated by the quadratic trial action, leading to quantitative disagreements with the PIMC results. This same physics underlies the distinct crossover observed in the phonon-mediated Coulomb screening of the exciton between 0 and 100 K (Fig.~\ref{Eb_int_FT}). As the temperature increases and phonon modes become thermally populated, the effective retarded interaction becomes better described by the quadratic approximation, progressively restoring agreement between the two methods. Moreover, since acoustic modes are generally quicker to thermally populate, the effects of polaron delocalization from thermal fluctuations are largely offset by the increased acoustic coupling.



\section{Conclusions}
We have parameterized an excitonic Hamiltonian from first principles methods to provide more predictive power for materials design. Specifically, we adapt this Hamiltonian to a path integral framework and integrate out the phonon degrees of freedom. Using PIMC, we were able to sample from the full non-perturbative partition function across different temperatures. We have demonstrated good agreement with experimental ground-state exciton binding energies for multiple materials with varying degrees of charge-phonon coupling strengths. We have also captured the interplay between the short-ranged deformations generated from acoustic modes and the long-ranged dipolar interactions generated from longitudinal optical modes. Furthermore, while we have demonstrated the utility of this model in a position representation for Wannier--Mott excitons composed of a single valence and conduction band, the model can be generalized by including the effects of lattice anharmonicity or working in a momentum representation.

The main advantage of a path-integral framework is that the exciton--phonon coupling can be captured non-perturbatively. To understand this, note that the partition function in Eqn.~\ref{Partition} captures all possible paths in the phase space spanned by the exciton and phonon degrees of freedom. As a result, the Fan-Migdal self-energy diagrams from MBPT\cite{fan1951temperature,migdal1958interaction,giustino2017electron} are resummed to all orders in the partition function.\cite{feynman1955slow}

From Eqn.~\ref{influence_general_compact}, we see that the non-local action contains a double integral over imaginary time with a phonon propagator that couples carrier positions at different imaginary times. When the carrier indices refer to the same quasiparticle, these terms generate attractive self-interactions. In the language of MBPT, this double time integral over the phonon propagator corresponds to a resummation of the Fan-Migdal self-energy diagrams to all orders. We note that our influence functional contains only first-order charge-phonon coupling vertices, and therefore does not capture the Debye-Waller contribution, which arises from a second-order coupling vertex in the MBPT framework.\cite{giustino2017electron}

When the carrier indices refer to different quasiparticles, Eqn.~\ref{influence_general_compact} generates repulsive phonon-mediated interactions between beads on the electron and hole ring polymers. These terms describe the dynamical screening of the electron--hole interaction by the lattice. In our formalism, the bare Coulomb interaction between the electron and hole is screened by the high-frequency dielectric constant, which accounts for the electronic contribution to the screening but contains no frequency dependence. The phonon-mediated repulsion between the electron and hole then provides the additional ionic contribution to the dielectric screening dynamically through the imaginary-time dependent phonon propagator. In the static limit, this screening would reduce the Coulomb interaction from its optically-screened value to one screened by the full static dielectric constant. However, because the phonon propagator retains its full imaginary time dependence, our framework captures the dynamical nature of this screening that is dependent on the relevant timescales of the exciton and the lattice.

The use of $\varepsilon_\infty$ for the Coulomb kernel is not merely a simplification but is required for consistency, since the ionic contribution to the screening is generated dynamically by the influence functional upon integrating out the phonon degrees of freedom. Using a frequency-dependent dielectric function in the Coulomb kernel would double-count the ionic screening at leading order, as the same lattice degrees of freedom would appear both in it and in the phonon-mediated interaction. We note, however, that the influence functional captures physics beyond what a frequency-dependent dielectric function in the Coulomb kernel would provide. Whereas screening enters through the electron--hole interaction at the level of linear response\cite{hybertsen1986electron} and is treated perturbatively in the exciton problem\cite{filip2021phonon,alvertis2024phonon}, the path-integral framework treats the carrier dynamics in the presence of the phonon-mediated interaction non-perturbatively, allowing the polaron formation and screening to self-consistently modify the excitonic state.

\section*{Acknowledgments}
Support for this work was provided by the U.S. Department of Energy, Office of Science, Office of Basic
Energy Sciences, Materials Sciences and Engineering Division, under Contract No. DEAC02-05-CH11231 within
the Fundamentals of Semiconductor Nanowire Program
(KCPY23). Computational resources were provided in
part by the National Energy Research Scientific Computing Center (NERSC), a U.S. Department of Energy Office of Science User Facility operated under contract no.
DEAC02-05CH11231. The authors thank Dr. Matt Coley-O'Rourke, Stephen Gant, and Daniel Chabeda for many useful discussions.




\bibliography{Cite}

\end{document}